\pdfoutput=1
\documentclass[a4paper,11pt]{article}
\usepackage[a4paper,left=2.73cm,right=2.7cm,top=3cm,bottom=3.5cm]{geometry}

\usepackage{hyperref}
\usepackage{amsmath,amsfonts,amssymb,amsthm,bm}
\usepackage{amsthm}
\usepackage{bbm}
\usepackage{color}
\usepackage{xcolor}
\definecolor{green}{RGB}{109, 163, 77}
\definecolor{violet}{RGB}{86 68 93}
\usepackage{tikz}
\usepackage{tikz-cd}
\usepackage{quiver}
\usepackage{graphicx}
\usepackage{cite}
\usepackage{comment}
\usepackage{cleveref}
\usepackage{framed}

\newtheorem{theorem}{Theorem}

\theoremstyle{definition}
\newtheorem{definition}{Definition}

\addtocontents{toc}{\protect\setcounter{tocdepth}{2}}
\numberwithin{equation}{section}

\def\Es{{\mathrm{E}_{7(7)}}}
\def\es{{\mathfrak{e}_{7(7)}}}
\def\Ed{{\mathrm{E}_{d(d)}}}

\def\SL{{\mathrm{SL}}}

\def\SU{{\mathrm{SU}}}
\def\SO{{\mathrm{SO}}}
\def\GL{{\mathrm{GL}}}
\def\Urm{{\mathrm{U}}}
\def\Mint{{M_{\text{int}}}}
\def\Mext{{M_{\text{ext}}}}
\def\Mtot{{M_{\text{tot}}}}

\def\Fr{{\text{Fr}}}

\newcommand{\spindle}{{\includegraphics[height=8pt]{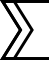}}}

\begin{document}

\begin{titlepage}

\thispagestyle{empty}

\begin{center}
{\LARGE \textbf{Consistent Truncations from Duality Symmetries\\ and Desingularization of Orbifold Uplifts}}

\vspace{40pt}

{\large \bf Anik Rudra}$\,^a$ \,  ,  \, {\large \bf Colin Sterckx}$\,^b$  \,  \large{and} \, {\large \bf Mario Trigiante}$\,^{c}$
 		
 \vspace{25pt}

$^a${\normalsize Mandelstam Institute for Theoretical Physics, School of Physics, University of the Witwatersrand, Private Bag 3, Wits 2050, South Africa} \\[7mm]


 $^b$\,{\normalsize
 Universit\'e Libre de Bruxelles (ULB) and International Solvay Institutes,\\
 Service  de Physique Th\'eorique et Math\'ematique, \\
 Campus de la Plaine, CP 231, B-1050, Brussels, Belgium.}
 \\[7mm]

 $^c$\,{\normalsize
 Department of Applied Science and Technology, Politecnico di Torino, Corso Duca degli Abruzzi 24,\\
 I-10129 Turin, Italy and INFN, Sezione di Torino, Italy.}
 \\[10mm]
 
\texttt{anikrudra23@gmail.com} \,\, , \,\,
  \texttt{colin.sterckx@ulb.ac.be}  \,\, , \,\, \texttt{mario.trigiante@polito.it}

\vspace{40pt}

\abstract{\noindent
This paper is an extension of the results presented in \cite{Guarino:2024gke}. We study $ G_S$-invariant subsectors of maximal gauged supergravities and show that such models can provide consistent truncations even when $G_S$ is not a symmetry of the original supergravity. We show that this construction is key to building pure supergravities around a supersymmetric AdS$_D$ solution. We illustrate this construction by building a consistent $\mathcal{N}=4$ subsector of the $D=4$ $\mathcal{N}=8$ $[\SO(6)\times \SO(1,1)]\ltimes \mathbb{R}^{12}$ gauged supergravity. We use this result to build the uplift of the multicharge spindle solutions in type IIB and we define a simple criterion for assessing the regularity of the uplift. We show that the type IIB uplift of the spindle is always non-regular, admitting eight codimension-six orbifold singularities. We apply the same criterion to other spindle uplifts, recovering known results and making predictions on the regularity of spindles on (quasi-)regular SE$_7$ manifolds.}
\end{center}
\end{titlepage}

\tableofcontents

\hrulefill
\vspace{10pt}

\section{Introduction}
\label{sec:Intro}

Lower-dimensional, gauged supergravities have provided a valuable framework for constructing and studying superstring/M-theory backgrounds of the form ${M}_{\text{ext}} \ltimes {M}_{\text{int}}\,,$
i.e. the (warped) product of an external $D$-dimensional spacetime manifold $M_{\text{ext}}$, and an internal, compact space, $M_{\mathrm{int}}$. On such backgrounds, it is often possible to consider a subsector of the full theory, reducing its full non-linear dynamics to that of a $D$-dimensional $\mathcal{N}$-extended gauged supergravity $SG[D,\mathcal{N}]$. The fluxes and the geometry of the internal manifold are encoded in the low-dimensional $SG[D,\mathcal{N}]$ as an \emph{embedding tensor}, specifying the gauge group $G_g$ and its embedding in the duality group $G_D$ \cite{Cordaro:1998tx,Nicolai:2000sc,deWit:2002vt,deWit:2005ub,deWit:2007kvg,Schon:2006kz,DallAgata:2023ahj} (see. \cite{Samtleben:2008pe,Trigiante:2016mnt,Inverso:2025zct} for reviews). When solutions in these truncated models uplift to solutions of the full equations of motion, we say that the truncation is \emph{consistent}. In this context, finding solutions for which ${M}_{\text{ext}}$ is maximally symmetric, reduces to the much simpler algebraic problem of extremising a scalar potential fully determined by the embedding tensor.

Exceptional Generalised Geometry (EGG) \cite{Coimbra:2011nw,Coimbra:2012af} and Exceptional Field Theory (ExFT) \cite{Hohm:2013vpa,Hohm:2013uia,Hohm:2014fxa}  have provided an ideal framework to describe type II and eleven-dimensional supergravities on ${M}_{\text{ext}}\ltimes{M}_{\text{int}}$ taking advantage of the hidden $\Ed$ structure of these theories (see \cite{Aldazabal:2013sca,Baguet:2015xha,Berman:2020tqn,Sterckx:2024vju,Samtleben:2025fta} for reviews). In this context, the appropriate mathematical tool to build consistent truncations to gauged supergravities is that of \emph{generalised G-structures}\cite{Coimbra:2011ky,Coimbra:2012af,Coimbra:2014uxa}. Within this framework, a lower-dimensional supergravity is a consistent truncation of a maximal ten or eleven-dimensional theory if it is the singlet sector of a structure group $G_S\subset \Ed$ and the corresponding $G_S$-structure is characterised by a $G_S$-singlet \emph{intrinsic torsion}\cite{Cassani:2019vcl,Josse:2025uro}. This general condition, based on EGG, has allowed for the embedding of non-maximally supersymmetric models within eleven-dimensional or type II theories, using a top-down approach based on general geometric properties of the internal manifold \cite{Gauntlett:2007ma,Larios:2019lxq,Cassani:2011fu}. 

In this language, a consistent truncation to a \emph{maximal} supergravity corresponds to an identity structure, satisfying the requirement of \emph{generalised parallelizability}. Hence, the generalised Scherk-Schwarz (gSS) reduction ansatz has provided a convenient, bottom-up method for embedding consistently maximally supersymmetric lower-dimensional models in eleven-dimensional or type II supergravities \cite{Hohm:2014qga}. For many purposes, however, maximal $D$-dimensional theories are too large (in four dimensions, they describe 70 scalar fields and 28 vector fields) and working with simpler, non-maximal models is desirable. The gSS ansatz can also be used to uplift such smaller models, provided they are consistently embedded in a maximal $D$-dimensional truncation. This brings about the main problem, to be dealt with in this paper, of building consistent subsectors of gauged supergravities, i.e. ``consistent truncations of consistent truncations".

A renowned example of consistent truncation is maximal supergravity in $D=4$ with gauge group $\mathcal{G}={\rm SO}(8)$, constructed in \cite{deWit:1982bul}, and describing a consistent truncation of eleven-dimensional supergravity on the Freund-Rubin solution ${\rm AdS}_4\times { S}^7$ \cite{deWit:1986oxb}. Several other vacua of the same four-dimensional theory were found and most of them were uplifted to variants of the Freund-Rubin background, of the form ${\rm AdS}_4\ltimes \tilde{ S}^7$, $\tilde{ S}^7$ being a 7-dimensional manifold with the same topology as the 7-sphere \cite{Warner:1983vz,Duff:1986hr,Comsa:2019rcz}. A systematic study of uplifts of maximal gauged supergravities was performed in \cite{Inverso:2017lrz,Inverso:2024xok}. Among the several other examples, here we shall be dealing with the maximal four-dimensional model with gauge group $\mathcal{G}=[\SO(6) \times \SO(1,\,1)]\ltimes \mathbb{R}^{12}$\cite{DallAgata:2011aa,Gallerati:2014xra,Inverso:2016eet}. Its anti-de Sitter vacua, which include a characteristic $\mathcal{N}=4$ one, are uplifted to a class of non-geometric backgrounds of type IIB superstring theory called \emph{J-folds}\cite{Inverso:2016eet,Bobev:2019jbi,Guarino:2020gfe,Giambrone:2021zvp,Bobev:2021yya,Giambrone:2021wsm,Bobev:2023bxs,Guarino:2024zgq}. These are characterised by a geometry of the form ${\rm AdS}_4\times S^1\times \tilde{ S}^5$, with a monodromy along $S^1$ in the ${\rm SL}(2,\mathbb{Z})$ global symmetry of the superstring theory, $\tilde{S}^5$ being a 5-dimensional manifold with the topology of $S^5$. We will refer to this gauged maximal supergravity as the \emph{J-fold model}.

Further examples of consistent truncations can be built by expanding around ${\rm AdS}_D\times M_{\mathrm{int}}$ solutions, preserving $\mathcal{N}'$ $D$-dimensional supersymmetries and provide consistent truncation to pure $\mathcal{N}'$-extended, $D$-dimensional supergravities. These truncations, whose consistency was conjectured in \cite{Gauntlett:2007ma} and later proven in \cite{Cassani:2019vcl} within EGG, are expected to encode the minimal gravitational dynamics holographically dual to conserved currents and stress-energy supermultiplets. These pure supergravity truncations are often quite useful in the context of holography as they can be used to build black hole, domain walls, instanton solutions, etc.

The conventional paradigm for constructing a consistent truncation of a gauged supergravity is to restrict to the singlet sector of a suitable compact symmetry group $G_0$ of the embedding tensor. Examples of consistent truncations of the maximal four-dimensional ${\rm SO}(8)$-gauged model are: the $\mathcal{N}'=2$ STU model with Fayet-Iliopoulos terms \cite{Duff:1999gh}, singlet sector of a ${\rm SO}(2)^3\subset {\rm SO}(8)$ group, or the pure $\mathcal{N}'=4$ supergravity, singlet sector of ${\rm SO}(4)\subset {\rm SO}(8)$\cite{Cvetic:1999au}. As proven in \cite{Guarino:2024gke}, however, the pure half-maximal $D=4$ supergravity containing the $\mathcal{N}'=4$ vacuum of the J-model cannot be built using the conventional paradigm, and this calls for a generalisation of the latter. 

Expanding on the results of \cite{Guarino:2024gke}, we will show that one can also consider consistent truncation with respect to the \emph{duality} group $G_S$ of a gauged supergravity, i.e. not necessarily an invariance of the embedding tensor, provided some constraints on the embedding tensor itself are satisfied. These conditions follow from the only requirement that the component of the embedding tensor, which is identified with the intrinsic torsion associated with a $G_S$-structure, be $G_S$-invariant. Far from being an esoteric construction, these consistent invariant subsectors allow one to capture the pure $\mathcal{N}'$ supergravities around any supersymmetric AdS$_D$ solutions. In particular, we will consider the $\mathcal{N}'=4$ vacuum of the J-fold maximal model and construct the corresponding truncation down to pure $\mathcal{N}'=4$ $D=4$ gauged supergravity. We will explicitly prove the consistency of this truncation at the level of the four-dimensional equations of motion and provide the explicit uplift ansatz of the pure $\mathcal{N}'=4$ $D=4$ gauged supergravity in type IIB supergravity. 
This completes the analysis of \cite{Guarino:2024gke}, in which only the explicit uplift formulae for an $\mathcal{N}=2$, $t$-model truncation of the pure $\mathcal{N}'=4$ supergravity were given.  Sufficient conditions for the consistency of a truncation of a maximal gauged supergravity in four dimensions were also recently derived, in a different form, in \cite{Pico:2025cmc}. 

As an application of our general paradigm, we shall discuss the uplift of the spindle solutions of this theory to type IIB backgrounds of J-fold type. In particular, we will study the uplift of the two-charge spindle solution of \cite{Ferrero:2021ovq} and prove the non-regularity of the corresponding ten-dimensional background. More specifically, building on the results of \cite{Rovere:2025jks}, we will provide a general criterion for the regularity of the uplifted spindle solutions which will allow us to recover various results in the literature, such as the regularity of single charge spindles in M-theory on regular SE$_7$ manifolds \cite{Ferrero:2020twa,Cassani:2021dwa}, the regularity of multi-charge spindles on $S^7$ \cite{Ferrero:2021ovq,Ferrero:2021etw,Couzens:2021rlk}, and the non-regularity, as well as the type of singularities, of seven-dimensional spindles \cite{Ferrero:2021wvk,Bomans:2024mrf}. 
Applying the same criterion to the spindle solutions, uplifted here to J-fold backgrounds in the type IIB theory, we shall prove the latter to be always singular, also in the supersymmetric cases, exhibiting eight codimension-six orbifold singularities. 

Finally, we will comment on various subsectors and $\mathcal{N}=2$ models as well as pure supergravity. Some comments about orbifolds are presented in Appendix \ref{app:orbifolds}.

\section{Gauged supergravities}

In this section, we will summarise the $\mathcal{N}=8$ and $\mathcal{N}=4$ $D=4$ gauged supergravities \cite{deWit:2007kvg,Schon:2006kz,DallAgata:2023ahj}. We focus on the two relevant theories for our work: the $\mathcal{N}=8$ with gauge group $[\SO(6) \times \SO(1,\,1)]\ltimes \mathbb{R}^{12}$ \cite{Inverso:2016eet}, and the $\mathcal{N}=4$ with gauge group $\SO(4)$, both containing a $\mathcal{N}=4$ AdS$_4$ solution \cite{Gallerati:2014xra,Louis:2014gxa}.

\subsection{The \texorpdfstring{$D=4$ $\mathcal{N}=8$}{D=4 N=8}  supergravities}
The $D=4$ $\mathcal{N}=8$ ungauged supergravity enjoys a global symmetry group $\Es$. Its bosonic field content comprises a metric, $g_{\mu\nu}$, 28 electric vector fields and their 28 magnetic duals, $A_\mu{}^M$, transforming in the $\mathbf{56}$ fundamental representation of $\Es$, as well as 70 scalars encoded in the coset representative $\mathcal{V} \in \Es/\SU(8)$. The gauged theory also contains a series of two-forms $B_\alpha$ transforming in the adjoint of $\Es$. The gauging procedure is performed by introducing an \emph{embedding tensor} $\Theta_M{}^\alpha$ transforming in the $\mathbf{912}$ of $\Es$ \cite{deWit:2007kvg}. Equivalently, this embedding tensor can be written in terms of $X_{MN}{}^P = \Theta_M{}^\alpha (t_\alpha)_M{}^N$, with $(t_\alpha)$ the generators of $\es$ in the $\mathbf{56}$ representation. For consistency, this embedding tensor must satisfy the \emph{quadratic constraints} (QC)
\begin{equation}
\begin{split}
    &\Theta_M{}^\alpha \Theta_N{}^\beta \Omega^{MN} = 0\,,\\
    &[X_M,\,X_N] = - X_{MN}{}^P X_P\,.
\end{split}
\end{equation}
It allows us to define a covariant derivative 
\begin{equation}
    D = d + A^M \Theta_M\,,
\end{equation}
and the field strengths
\begin{equation}
\begin{array}{ll}
    &\mathcal{H}^M  = dA^M + \frac{1}{2} A^N A^P X_{MN}{}^P - \frac{1}{2}\Omega^{MN}\Theta_N{}^\alpha B_\alpha\,,\\
    & \mathcal{H}_\alpha = DB_\alpha - t_{\alpha PQ} A^P (dA^Q + \frac{1}{3} X_{RS}{}^Q A^R A^S)\,.
\end{array}
\end{equation}
invariant under 0-form gauge transformations
\begin{equation}
    \begin{array}{l}
         \delta A_\mu^M = D_\mu \zeta^M \,, \\
          \delta B_\alpha = t_{\alpha NP} (2 \zeta^N  \mathcal{H}^P - A^N \delta A^P)\,,
    \end{array}
\end{equation}
and the 1-form gauge transformations
\begin{equation}
    \begin{array}{l}
         \delta_\Xi A_\mu^M =  -\frac{1}{2} \Omega^{MN} \Theta_N{}^\alpha \Xi_\alpha\,, \\
          \delta_\Xi B_\alpha = D \Xi_\alpha - t_{\alpha NP}A^N \,\delta_\Xi A^P\,.
    \end{array}
\end{equation}
The gauging procedure fixes completely the scalar potential of the theory. It is more easily expressed in terms of the $T$-tensor or $\Xi$-tensor:
\begin{equation}
    T_{MN}{}^P(\mathcal{V},\,X) = (\mathcal{V}^{-1} \star X)_{MN}{}^P = \Xi_M{}^\alpha (t_\alpha)_M{}^N
\end{equation}
\begin{equation}
    V= \frac{1}{672} \left[\text{Tr} (T_M T_M^T) + 7\, \text{Tr}(T_M T_M)\right]\,.
\end{equation}
or
\begin{equation}
    V(\Xi) = \frac{1}{56}\Xi_M{}^\alpha \Xi_M{}^\beta \left(\delta_{\alpha\beta} + \frac{7}{12}\, k_{\alpha\beta}\right)\,.
\end{equation}
for generators normalised as $\text{Tr}(t_\alpha t_\beta^T) = 12 \delta_{\alpha\beta}$ and $k_{\alpha\beta}$ is the Cartan-Killing metric. Since, under the joint action of $\Es$ on both the scalars and the embedding tensor, the $T$-tensor only transforms under the ${\rm SU}(8)$ compensating transformation, the scalar potential, in particular, as well as the whole field equations, are invariant under such combined  $\Es$-action. This proves that  $\Es$ is a duality group of maximal gauged supergravities.

\paragraph{Fermions}
The maximal supergravity contains also eight gravitini $\psi_\mu{}^A$ and 56 dilatini, $\chi_{[ABC]}$ transforming under the $\SU(8)_R$ R-symmetry group of the theory. Although the precise coupling of the fermions to the rest of the theory is not particularly useful for our discussion, we recall the definitions of the fermion-shift matrices. These matrices encode the amount of preserved supersymmetries in a given AdS$_4$ vacuum. One has
\begin{equation}
    \begin{split}
    &(A_1)^{AB} = \frac{\sqrt{2}}{21} T^{AEFB}{}_{FE}\,,\\
    &(A_2)_{A}{}^{BCD} = \frac{1}{\sqrt{2}} T_{AE}{}^{EBCD}\,.
    \end{split}
\end{equation}
The scalar potential can be written in terms of those matrices as
\begin{equation}
    V_{\mathcal{N}=8} = \frac{1}{24} |A^A{}_{BCD}|^2 - \frac{3}{4} |A_{AB}|^2\,.
\end{equation}
When $\mathcal{N}'$ supersymmetries are preserved at a given extremum of the scalar potential given by the scalars $\phi_0$, we can split the $\SU(8)$ index $A$ into $\alpha = 1 ,\,\dots,\,\mathcal{N}'$ and $a = \mathcal{N}'+1,\,\dots,8$, and, up to a global $\SU(8)$ transformation, $(A_2)_\alpha =0$, $(A A^\dagger)_\alpha{}^\beta = -\frac{1}{6} V(\phi_0) \delta_\alpha{}^\beta$, and $(A A^\dagger)_a{}^{\alpha} = 0$.

\paragraph{The $[\SO(1,1)\times \SO(6)]\ltimes\mathbb{R}^{12}$ gauging and the $\mathcal{N}=4$ vacuum}

The specific embedding tensor is given, in $\SL(8) \subset \Es$ representation as
\begin{equation}
    \Theta_{[AB]}{}^C{}_D = 4\sqrt{3} \delta_{[A}^C\theta_{B]D} \hspace{1cm} \Theta^{[AB]C}{}_D = 4\sqrt{3} \delta^{[A}_D\tilde{\theta}^{B]C}\,,
\end{equation}
where $\theta = g \,\text{diag}(0,\,1_6,\,0)$ and $\tilde{\theta}=c\,\text{diag} (-1,\,0_6,\,1)$. The $\mathcal{N}=4$ vacuum is obtained when the scalar representative takes the value
\begin{equation}
\begin{split}
    \mathcal{V}_{\mathcal{N}=4} =& \text{exp}\left[(t_{1234}+t_{1567})\right]\\
    &\cdot \text{exp}\left[- \frac{\ln 2}{2} (t_1{}^1-t_{8}{}^8) - \frac{\ln |c|}{4} (t_2{}^2 + \cdots t_{7}{}^7 - 3 t_1{}^1 - 3 t_8{}^8) \right]\,.
\end{split}
\end{equation}
These $\es$ generators are normalised to $\text{Tr}_{56}(t \cdot t^T) = 12$.

All the vacua of this model, including the aforementioned half-maximal one, uplift to non-geometric backgrounds of type IIB supergravity known as J-folds \cite{Inverso:2016eet}, we shall refer to this gauged maximal supergravity as the \emph{J-fold model}. We note that this $\mathcal{N}=4$ vacuum admits two exactly marginal deformations retaining eight supercharges and whose parameters $\chi,\,\delta$ are dual to marginal deformations of the J-fold model at the boundary, see \cite{Guarino:2020gfe,Giambrone:2021zvp,Guarino:2021kyp,Arav:2021gra,Bobev:2021yya,Guarino:2021hrc,Cesaro:2021tna}.\footnote{In fact vacua marginally connected to the $\mathcal{N}=4$ form a three-dimensional manifold of generically non-supersymmetric solutions \cite{Giambrone:2021wsm,Bobev:2023bxs}. } The half-maximal vacuum corresponds to the point $\chi=0,\,\delta=1$. For the interpretation of $\chi$, as a twist of the internal $S^5$ over the internal $S^1$, see \cite{Giambrone:2021zvp,Guarino:2021hrc}, while an analogous interpretation of $\delta$ is still elusive.

\subsection{Pure \texorpdfstring{$D=4\,\mathcal{N} = 4$}{D=4 N=4} SO(4) gauged supergravity}
\label{sec:D=4N=4}

 Let us review the pure $\mathcal{N}=4$ theory which admits an $\SO(6)\times \SL(2)$ duality group. The gravity multiplet contains the graviton $g_{\mu\nu}$, four gravitini $\psi_\mu^i$, six vectors $A_\mu^{m +}$ and their magnetic duals $A_\mu^{m -}$ (written together as $A_\mu^{m\alpha}$) and transforming as the $(\mathbf{6},\,\mathbf{2})$ under the duality group, four spin-$1/2$ fermions $\chi^i$ and two scalars parametrising the $\SL(2)/\SO(2)$ coset space. The embedding tensor is given by two tensors $f_{\alpha\,mnp} \in (\mathbf{20},\,\mathbf{2})$ and $\xi_{\alpha \,m} \in (\mathbf{6},\,\mathbf{2})$. In order for the theory to admit a $\mathcal{N}=4$ solution\cite{Louis:2014gxa}, we must fix $\xi_{\alpha \,m} = 0$. Concerning $f_{\alpha\,mnp}$, several equivalent embedding tensors can be chosen. To describe our chosen embedding, we split the index $m = a,\,\hat{a}$ with $a=1,\,2,\,3$ and $\hat{a}=4,\,5,\,6$ and we fix
\begin{equation}
    f_{+\,{123}} = 1 \hspace{1cm} f_{-{456}} = 1\,.
\end{equation}
equivalently $f_+ = \epsilon_{abc}$ and $f_-=\epsilon_{\hat{a}\hat{b}\hat{c}}$.
This choice fixes the gauge group to be
\begin{equation}
    G_{g} = \SO(3)_{+} \times \SO(3)_{-} \subset \SO(6)_R\,.
\end{equation}
The scalar coset space is parametrised by the symmetric matrix
\begin{equation}
    \mathcal{M} = -\begin{pmatrix}
         e^\xi & 0 & -\chi\,e^\xi  & 0 \\
        0  & e^\xi& 0 &-\chi\,e^\xi \\
       -\chi\,e^\xi & 0 & e^\xi|\tau|^2 & 0 \\
        0 &-\chi\,e^\xi & 0 & e^\xi |\tau|^2\\
    \end{pmatrix} \otimes \mathbbm{1}_{3\times 3}\,.
\end{equation}
We define the gauge-invariant improved field strengths for the electric vectors $A^{a+}$ and $A^{\hat{a}-}$
\begin{equation}
\begin{split}
    \mathcal{H}^{a+} &= dA^{a+} + \frac{1}{2} \epsilon_{bc}{}^{a}A^{b+}A^{c+}\,,\\
    \mathcal{H}^{\hat{a}+}& = dA^{\hat{a}+} +\frac{1}{2}\epsilon_{\hat{b}\hat{c}}{}^{\hat{a}} A^{\hat{b}-} A^{\hat{c}+}+\frac{1}{2} B^{\hat{b}\hat{c}}\epsilon_{\hat{b}\hat{c}}{}^{\hat{a}}\,,\\
    \mathcal{H}^{a-}& = dA^{a-} +\frac{1}{2}\epsilon_{bc}{}^{a} A^{b+} A^{c-}+ \frac{1}{2} B^{bc}\epsilon_{bc}{}^{a}\,,\\
    \mathcal{H}^{\hat{a}-}&= dA^{\hat{a}-} + \frac{1}{2} \epsilon_{\hat{b}\hat{c}}{}^{\hat{a}}A^{\hat{b}-}A^{\hat{c}-}\,.
\end{split}
\end{equation}

With our gauging, the bosonic action of \cite{Schon:2006kz} reduces to
\begin{equation}
    \mathcal{L}_{\text{bos}} = \mathcal{L}_{\text{kin}} + \mathcal{L}_{\text{top}} + \mathcal{L}_{\text{pot}}\,.
\end{equation}

\begin{align}
    e^{-1} \mathcal{L}_{\text{kin}} =& \tfrac{1}{2} R - \frac{ \partial_\mu \tau \partial^\mu \tau}{4 \,\text{Im}(\tau)^2}\\
    & - \frac{1}{4} \text{Im}(\tau) \delta_{mn} \mathcal{H}^{m+}_{\mu\nu} \,\mathcal{H}^{\mu\nu\,n^+} + \frac{1}{8} \text{Re}(\tau) \delta_{mn} \epsilon^{\mu\nu\rho\sigma} \mathcal{H}^{m+}_{\mu\nu}\mathcal{H}_{\rho\sigma}^{n+}\,.\\[2mm]
    e^{-1}\mathcal{L}_{\text{top}} =& - \frac{g}{2} \epsilon^{\mu\nu\rho\sigma} \Big( f_{-\,mnp} A^{m-}_\mu A^{n+}_\nu \partial_\rho A_\lambda^{p-} - \frac{g}{4} \delta_{\alpha\beta} A_\mu^{[p|\alpha|} A_\nu^{q]+} A_\rho^{[p|\alpha|} A_\lambda^{q]-}\\
    &\hspace{15mm} - \frac{1}{4} f_{-mnp} B^{np}_{\mu\nu} \left( \partial_\rho A^{m-}_\lambda - g {f_{\alpha qr}}^m A_\rho^{q\alpha} A_{\lambda}^{r-}\right)\Big)\\[2mm]
   e^{-1} \mathcal{L}_{\text{pot}} =&  \frac{1}{6} \text{Tr}\left(M^{\alpha\beta}\right) -2 \,.
\end{align}
The four-dimensional equations of motion read
\begin{equation}
\begin{split}
    &\mathcal{H}^{[mn]}= 0\,,\\
    &\square \xi - e^{2\xi} \partial_\mu \chi \partial^\mu \chi - \partial_\xi V + \mathcal{H} \partial_\xi\mathcal{M} \mathcal{H} = 0\,,\\
    &\nabla_\mu\left(e^{2\xi} \partial^\mu \chi\right) - \partial_\chi V + \mathcal{H} \partial_\chi\mathcal{M} \mathcal{H} =0\,,
\end{split}
\end{equation}
whereas the twisted self-duality condition gives
\begin{equation}
   \star \mathcal{H}^{m\alpha}= - (\mathbb{C} \mathcal{M} \mathcal{H})^{m\alpha}\,.
   \label{eq:selfdualityeqN4}
\end{equation}
It allows us to read the dual field strengths from the usual electric ones as
\begin{equation}
    \begin{split}
        &\mathcal{H}^{a-} = |\tau|^{-2} \left(e^{-\xi} \star \mathcal{H}^{a+} + \chi \,\mathcal{H}^{a+}\right)\,,\\
        &\mathcal{H}^{\hat{a}+} = - e^{-\xi} \star \mathcal{H}^{\hat{a}-} + \chi \mathcal{H}^{\hat{a}-}\,.
    \end{split}
\end{equation}
Using these formulae, we can remove the two-forms from the equations of motion and consider $(\mathcal{H}^{a+},\,\mathcal{H}^{\hat{a}-})$ as the two independent field strengths in the uplift formulae we will provide later on for the $p$-form field strengths in type IIB.

\section{On consistent truncations}\label{sec:3}

For illustration, consider an AdS$_4$ $\mathcal{N}'$-supersymmetric solution of a four-dimensional $\mathcal{N}=8$ gauged supergravity. One may wish to isolate the degrees of freedom dual to the stress-energy tensor of the corresponding SCFT$_3$. These are expected to coincide with the fields of a pure $\mathcal{N}'$ supergravity, and the associated subsector should therefore be identified with the $G_S = \mathrm{SU}(8-\mathcal{N}') \subset \mathrm{SU}(8)_R$ invariant sector of the parent theory. However, in the presence of a gauging, the full theory need not be invariant under $\mathrm{SU}(8-\mathcal{N}')$, so the standard symmetry-based arguments ensuring consistency of such a truncation do not directly apply. More generally, we will consider gauged supergravities with duality group $G_D$ and show that certain $G_S$-invariant subsectors, with $G_S \subset G_D$, define consistent truncations even when $G_S$ is not a symmetry of the theory -- namely, when it does not leave the embedding tensor invariant. Following the work of \cite{Cassani:2019vcl}, we will show that consistency only requires that specific components of the embedding tensor, referred to as “intrinsic,” be $G_S$-invariant. 

In this section, we provide a general proof of this statement, based on generalized geometry, which applies whenever the parent theory admits an uplift. In addition, for supersymmetric truncations, we demonstrate directly within the lower-dimensional supergravity that the resulting subsectors indeed define consistent truncations. This is completed by the results in appendix \ref{App:susyCT}, where we will show that around a supersymmetric vacuum, $\SU(8-\mathcal{N}')$-invariant subsectors are always consistent. This establishes a lower-dimensional analogue of the Gauntlett–Varela conjecture \cite{Gauntlett:2007ma}.

\subsection{Systematics of consistent truncations in standard geometry}

Following the presentation of \cite{Cassani:2019vcl}, we first review the concepts of $G$-structure and intrinsic torsion in the context of ordinary (i.e. non-generalised) geometry. These notions are necessary to understand various properties of consistent truncations and how to build them in light of theorem \ref{thm:CTrStandard}.

\subsubsection{$G$-structures and consistent truncations}

In the context of dimensional reductions, one considers the total manifold as a warped product of an ``external" manifold, $\Mext$, of dimension $D$, and an ``internal" manifold, $\Mint$, of dimension $d$. In this context, the following statement holds:
\begin{theorem}
    General relativity defined on a product manifold $\Mint \times \Mext$ admits a consistent truncation to a $D$-dimensional theory of gravity on $\Mext$ coupled to gauge vectors and scalar fields whenever $\Mint $ admits a $G_S$-structure with $G_S$-invariant and constant intrinsic torsion. This truncation is built by expanding all metric fields as $G_S$-invariant tensors.
    \label{thm:CTrStandard}
\end{theorem}
\noindent Let us unpack the various definitions present in this theorem and spell out the construction of the consistent truncation.

The reduction of the structure group of a manifold is defined as a reduction of its frame bundle $\Fr(\Mint)$\footnote{I.e. it is a $G_S$-principal bundle on $\Mint$, $P_{G_S}$, such that $\Fr(\Mint) \cong P_{G_S}\times_{G_S} \GL(d)$.}. This simply signals that all the non-trivial information on $\text{Fr}(\Mint)$ is actually contained in a smaller principal bundle $P_{G_S}$ on $\Mint$. This definition is not particularly easy to work with. For our purpose, it will be sufficient to consider the more restrictive definition:
\begin{definition}[$G_S$-structure from sections] A $G_S$-structure on a manifold $\Mint$ is a subgroup $G_S$ of $\GL(d)$ and a set of tensors on $\Mint$: $I_{G_S} =\{\Xi_i| i\in I\}\,,$ such that $\text{Stab}\left(\Xi_i(p)\right)\cong G_S$ $\forall\, p \in \Mint$.
\end{definition}
\noindent There is another useful, and equivalent, definition of a $G_S$-structure in terms of frames:
\begin{definition}[$G_S$-structure from a frame] A $G_S$-structure on a manifold $\Mint$ of dimension $d$ is given by a subgroup $G_S$ of $\GL(d)$ and an element $e \in \Gamma(G_S\backslash\Fr(\Mint))$, invertible at any point on $\Mint$.
\end{definition}
\noindent In a chart, the frame is just a vielbein $e^a{}_m dx^m$ defined up to the $G_S$ action on the index $a$.

\paragraph{Compatible connections and torsion}
Given a reduction of the structure group on $\Mint$, we say that a connection $\nabla$ on $\Fr(\Mint)$ is \emph{compatible} if
\begin{equation}
    \nabla \Xi = 0 \hspace{5mm},\hspace{5mm}\forall\, \Xi\in I_{G_S}\,.
\end{equation}
It is the connections leaving invariant the $G_S$-invariant tensors. The set of compatible connections is always non-empty and modelled, as an affine vector space, on $ K_{G_S}=T^*M \otimes \mathfrak{g}_S$. Indeed, for two compatible connections $\nabla$ and $\nabla'$, $\nabla-\nabla'$ is a tensor acting trivially on $G_S$-invariant tensor thus $\nabla - \nabla' \in \Gamma(T^*M \otimes \mathfrak{g}_S)$. We define the torsion of a connection as
\begin{equation}
\begin{split}
    T^\nabla: \Lambda^2T\Mint \rightarrow T\Mint:(X,\,Y)\mapsto&\nabla_X Y - \nabla_Y X - [X,\,Y] \\
    &= \mathcal{L}^\nabla_X Y - L_X Y\,, 
\end{split}
\end{equation}
where $\mathcal{L}$ is the Lie derivative and $\mathcal{L}^\nabla$ is the covariantised Lie derivative. Equivalently, the torsion tensor can be understood as a section of a subspace of $T^*M \otimes \mathfrak{gl}_d$ given by $\mathcal{L}^\nabla - \mathcal{L}$. We will denote the space of torsions:
\begin{equation}
    W = \{T^\nabla \,|\, \nabla \text{ a connection on }\Mint\}.
\end{equation}
Notice that $T^\nabla$ does depend on the choice of compatible connection and is not an intrinsic property of the reduction of the structure group. To cure this, consider two compatible connections $\nabla$ and $\nabla' = \nabla + \Omega$ with $\Omega \in K_{G_S}$, and consider the map
\begin{equation}
    \tau_{G_S}: K_{G_S} \rightarrow W : \Omega \mapsto T^\nabla - T^{\nabla- \Omega}\,.
\end{equation}
This map provides an embedding of $K_{G_S}$ in the space of torsions and we define $W_{G_S} = \text{Im}(\tau_{G_S}) \subset W$.
\begin{definition}
    Given a $G_S$ structure, the \emph{space of intrinsic torsions} is the vector space 
    \begin{equation}
    W^{int}_{G_S} = W / W_{G_S}\,,
\end{equation}
and we denote $\pi_{G_S} : W \rightarrow W^{int}_{G_S}$ the associated projector.
The \emph{intrinsic torsion} of the $G_S$-structure is 
\begin{equation}
    T^{int} = \pi_{G_S}\left(T^\nabla\right)
\end{equation}
for any compatible connection $\nabla$. The intrinsic torsion is, by construction, independent of the choice of compatible connection $\nabla$. 
\end{definition}

\paragraph{The $G_S$-invariant subsector and consistency}
Having defined all relevant objects, we can build a consistent truncation from them. First, the field content of the $G_S$-invariant subsector is spanned by scalars representing $G_S$-invariant degrees of freedom of the internal metric. They parametrise the coset space
\begin{equation}
    \mathcal{M}_{\text{scal}} = \frac{\text{Comm}_{\GL(d)}(G_S)}{\text{Comm}_{\textrm{O}(d)}(G_S)}\,.
\end{equation}
The Kaluza-Klein vectors, originating from the off-diagonal terms of the metric, are obtained from the invariant one-forms $\eta_a$ in $I_{G_S}$:
\begin{equation}
    A^{KK} = A(x^\mu)^a \eta_a\,.
\end{equation}
Finally, the truncated theory also contains the degrees of freedom originating from the external metric $g_{\mu\nu}$.

To show that this subsector is consistent, we must show that the equations of motion evaluated at a $G_S$-invariant point are themselves $G_S$-invariant. Obviously, any algebraic combination of $G_S$-invariant modes will still be $G_S$-invariant. However, one should worry about terms of the form $\nabla^{L.C.} \Xi$ might not be $G_S$-invariant. Since the Levi-Civita connection is torsionless, we have that
\begin{equation}
    \nabla^{L.C.} \Xi = \nabla \Xi - T\cdot \Xi  \in I_{G_S}
\end{equation}
where $\nabla$ is a compatible connection and $T\cdot$ is the action of the intrinsic torsion on the invariant tensors. This shows that consistency only requires that the \emph{intrinsic} torsion is a constant $G_S$-invariant tensor.

\subsubsection{Generalised $G_S$-structures}

The constructions we reviewed in the context of standard geometry can be extended to generalised geometry as was done in \cite{Coimbra:2011ky,Coimbra:2012af,Coimbra:2014uxa}. In exceptional generalised geometry, the structure group $\GL(d)$ is replaced by one of the exceptional Lie group $\Ed\times \mathbb{R}^+$. Generalised vectors are sections of a generalised vector bundle $E$, which is an associated bundle in specific representations of $\Ed$. We will also denote by $F$ the vector bundle associated with the adjoint representation of $\Ed \times \mathbb{R}^+$. Once again, a generalised $G_S$-structure is defined by a set of $G_S$-invariant tensors $I_{G_S} = \{\Xi_i\}$. A connection $\nabla$ on the generalised tangent bundle is compatible if 
\begin{equation}
    \nabla \Xi=0\hspace{5mm}\hspace{5mm}\forall\, \Xi\in I_{G_S}\,.
\end{equation}
The torsion of such a connection is obtained by considering the tensor
\begin{equation}
    T^\nabla = L^\nabla - L \in \Gamma(E^* \otimes F)\,,
\end{equation}
where $L$ is the generalised Lie derivative:
\begin{equation}
	L_\Lambda V^M = \Lambda^K\partial_K V^M - \alpha_d\, \mathbb{P}_{\text{adj.}}{}^{M}{}_N{}^P{}_Q\, \partial_P \Lambda^Q\, V^N \,,
	\label{eq:genLieDer7}
\end{equation}
where $\Lambda$ and $V$ $\in\Gamma(E)$ are generalised vectors, $\alpha_d$ is a real number depending on $d$, and $\mathbb{P}$ is the projector on the adjoint representation. The space of torsions only spans a subspace $\Gamma(W)$ of $\Gamma(E^* \otimes F)$.\footnote{For example, in the $\Es$ case $T \in \mathbf{56} \oplus \mathbf{912}\oplus \mathbf{6480}$ but the generalised Lie derivative projects out the $\mathbf{6840}$ representation.}

As before, we can define a notion of intrinsic torsion associated with the reduction of a structure group. The space of compatible connections is still modelled on $K_{G_S}=E^* \otimes \mathfrak{g}_S$ and for any compatible connection $\nabla$, we define the embedding
\begin{equation}
    \tau_{G_S} : K_{G_S} \rightarrow W: \Omega \mapsto T^{\nabla} - T^{\nabla-\Omega}\,.
\end{equation}
We define $W_{G_S} = \text{Im}(\tau_{G_S})$ and the space of intrinsic torsion is
\begin{equation}
    W_{G_S}^{int} = W/W_{G_S}\,.
\end{equation}
As before, any reduction of the structure group with constant singlet intrinsic torsion, provides a consistent truncation to a gauged supergravity. The reduced gauged supergravity will admit $\mathcal{N}'$ supersymmetric transformations, where $\mathcal{N}'$ is the number of gravitini invariant under $G_S$. Its duality group will be 
\begin{equation}
    G_D = \text{Comm}_{\Ed}(G_S)\,,
\end{equation}
and the scalar target space will read
\begin{equation}
    \mathcal{M}_{\text{scal}} = \frac{\text{Comm}_{\Ed}(G_S)}{\text{Comm}_{\Ed}(H_d)}\,,
\end{equation}
where $H_d$ is the maximally compact subgroup of $\Ed$. It will contain vectors spanning the $G_S$-invariant subspace of $E$ and its gauging will be specified by the intrinsic torsion, itself identified with the embedding tensor of the lower-dimensional theory. We refer to \cite{Josse:2025uro} for an illustrative list of examples of $G_S$ and their corresponding supergravity theory.

\subsection{Consistent truncations of consistent truncations}
\label{sec:ctsquared}
Let us consider a consistent truncation obtained from a generalised $G_S$-structure whose intrinsic torsion $T_{G_S}$ is a constant $G_S$-singlet tensor.
Consider a further truncation to a $G_S' \subset \text{Comm}_{\Ed}(G_S)$ invariant subsector. When the embedding tensor is $G_S'$-invariant, by the usual symmetry argument, this subsector is a consistent truncation of the full theory. This is the standard case where $G_S'$ is a subgroup of the gauge and the global symmetry group. We are interested in the case where $G_S'$ does not act trivially on the embedding tensor.

The reductions of the structure groups are defined in terms of  sets of invariant tensors 
\begin{equation}
    I_{G_S' \times G_S} \subset I_{G_S}\,.
\end{equation}
 Since $I_{G_S' \times G_S}$ is a subset of $I_{G_S}$, a compatible connection for the $G_S$ structure is a compatible connection for the $G_S' \times G_S$ structure. Moreover, $K_{G_S \times G_S'} = K_{G_S}+ K_{G_S'}$ giving us the series of projections 
\begin{equation}
\begin{array}{rll}
    W &\stackrel{\pi_{G_S}}{\longrightarrow} W^{int}_{G_S}=W/W_{G_S}&\stackrel{\pi_{G_S'}}{\longrightarrow}W^{int}_{G_S \times G_S'} = W/(W_{G_S} + W_{G_S'}).
\end{array}
\end{equation}
This allows us to state the following criterion for consistency of the $G_S'$-invariant subsector:
\begin{framed}
A $G_S'$-invariant subsector is consistent whenever the \emph{intrinsic} torsion given by $\pi_{G_S'}(T_{G_S})$ is $G_S'$-invariant. This does \emph{not} require $T_{G_S}$ itself to be $G_S'$-invariant. It only implies that the non-invariant pieces of $T_{G_S}$ are all projected out by $\pi_{G'_S}$, i.e. they are all \emph{non-intrinsic}.
\end{framed}

 To evaluate whether or not this property is satisfied for a given embedding tensor, one studies the branching $\Es \rightarrow G_D \times G_S' \times G_S$. If the non-$G_S$'-invariant pieces of the embedding tensors are all in representations of $W_{G_S} + W_{G_S'}$, then the truncation will be consistent. In our analysis below, we shall refer to the non-$G_S'$-invariant components of $W^{int}_{G_S \times G_S'}$ by $W^{(bad)}$ as this space contains all the components of the embedding tensor obstructing the consistency of the truncation provided by the $G_S'$-invariant sector. In section 4, we will provide a detailed realisation of such an exotic consistent truncation when $G_S = e$ and $G_S' = \SU(4)_S$.

\paragraph{Linear and quadratic constraints}
For practical computations, we rephrase our conditions for consistency in terms of explicit tensors. Although our claim is general, in the discussion below, for the sake of concreteness, we shall restrict to the case in which the parent theory is a maximal supergravity $G_S=\{e\}$ and, for notational simplicity, denote $G'_S$ simply by $G_S$. We will split the relevant representations of $\Ed$ in the following way: the fundamental representation index $M \rightarrow I \oplus \hat{I}$, where $I$ are the $G_S$-invariant indices, and the adjoint representation index splits as $\alpha\rightarrow \alpha_S \oplus \alpha_D \oplus \alpha_m$, where $\alpha_S$ labels generators of $\mathfrak{g}_S$, $\alpha_D$ labels the generators of $\mathfrak{g}_D$, and $\alpha_m$ label generator transforming in ``mixed" representation of $\mathfrak{g}_S$ and $\mathfrak{g}_D$ (i.e. which are neither singlets of $G_S$ nor $G_D$).

With these conventions, when the intrinsic torsion of a $G_S$-structure is a constant $G_S$-singlet, any of its representatives are of the form $T = \theta + \tau_{G_S}(\Omega)$ for $\Omega \in E \otimes \mathfrak{g}_S$ and $\theta$ $G_S$-invariant. We compute that
\begin{equation}
\begin{split}
    \tau_{G_S}(\Omega)_\Lambda(V) &= \Lambda^M(T^{\nabla} - T^{\nabla-\Omega})_{M}{}^\alpha V_\alpha\\ &= L_\Lambda^{\nabla} V - L_\Lambda^{\nabla-\Omega} V\\
    &=\Lambda^M (\Omega_M{}^\alpha - \alpha_d (t_\beta t^\alpha)_M{}^R \Omega_R{}^\beta) (t_\alpha)_P{}^N V^P\\
    &=: \Lambda^M \tau(\Omega)_M{}^\alpha (t_\alpha)_P{}^N V^P
\end{split}
\end{equation}
Then, the full torsion can be written as 
\begin{equation}
\begin{split}
    &T_I = \theta_I{}^{\alpha_D} (t_{\alpha_D}) + \Omega_I{}^{\alpha_S} (t_{\alpha_S})\\
    &T_{\hat{I}} = \theta_{\hat{I}}{}^{\alpha_m} (t_{\alpha_m}) + \theta_{\hat{I}}{}^{\alpha_s} (t_{\alpha_s})+ \Omega_{\hat{I}}{}^{\alpha_S}(t_{\alpha_S})\\
    &\hspace{3.8cm}\phantom{T_{\hat{I}} =}-\alpha_d \left[(t_{\beta_S})_{\hat{I}}{}^{\hat{J}} (t^{\alpha_s})_{\hat{J}}{}^{\hat{K}} \Omega_{\hat{K}}{}^{\beta_S}  \right](t_{\alpha_S}) \\
    &\hspace{3.8cm}\phantom{T_{\hat{I}} =}- \alpha_d\left[(t_{\beta_S})_{\hat{I}}{}^{\hat{J}}(t^{\alpha_m})_{\hat{J}}{}^M \Omega_M{}^{\beta_S} \right] (t_{\alpha_m})\,.
    \label{eq:embTensorSplit}
\end{split}
\end{equation}
Since $\theta$ is $G_S$-invariant, this computation shows that
\begin{equation}
    T_I{}^{\alpha_m} =0 \hspace{5mm}\text{ and }\hspace{5mm} T_{\hat{I}}{}^{\alpha_D} = 0\,.
\end{equation}
Compatible conditions were given in \cite{Pico:2025cmc}. The tensor $\theta_I{}^{\alpha_D}$ can be identified with the embedding tensor of the reduced theory. In particular, since we only turn on $G_S$-singlet in the reduced theory, the full covariant derivative $D = d + A^M \theta_M$, reduces to $D = d + A^I \theta_I{}^{\alpha_D} (t_{\alpha_D})$, which is the covariant derivative of the reduced theory.

Using this, we can consider the quadratic constraints of the subsector, assuming that those of the full theory are valid i.e.
\begin{equation}
\begin{split}
    &\Omega^{MN} T_M T_N = 0\,,\\
    &[T_M,\,T_N] = - T_{MN}{}^P T_P\,.
\end{split}
\end{equation}
Splitting those in the various $G_S$ and $G_D$ irreducible representations, we want to highlight the following quadratic constraint:
\begin{equation}
    \begin{split}
    &\Omega^{MN} T_M{}^{\alpha_D}T_N{}^{\beta_D} = \Omega^{IJ} \theta_I{}^{\alpha_D} \theta_J{}^{\beta_D} = 0\,,\\
    &\Omega^{MN}  T_M{}^{\alpha_D}T_N{}^{\beta_s} = \Omega^{IJ} \theta_I{}^{\alpha_D} \Omega_{J}{}^{\beta_S} + \underbrace{\Omega^{I\hat{J}}}_{=0}(\cdots)_{I\hat{J}} = 0\,.
    \end{split}
\end{equation}
In the first line we recognise part of the QC of the reduced theory. Moreover, we can compute
\begin{equation}
\begin{split}
    [T_I,\,T_J] &= - T_{IJ}{}^M T_M \,\Rightarrow\,[\theta_I,\,\theta_J] + [\Omega_I,\,\Omega_J]= - \theta_{IJ}{}^K (\theta_K+\Omega_K) 
\end{split}
\end{equation}
Reading the $\mathfrak{g}_D$ and $\mathfrak{g}_S$ components we get
\begin{equation}
\begin{split}
     &[\theta_I,\,\theta_J] = - \theta_{IJ}{}^K \theta_K\\
     &[\Omega_I,\,\Omega_J] = - \theta_{IJ}{}^K \Omega_K
\end{split}
\end{equation}
The first line is the usual QC for the reduced theory, while the second line is an extra constraint on $\Omega_I$ implying that it closes as a subalgebra of $G_g$, i.e. $\Omega_I$ induces a representation (possibly non faithful) of $G_g$ in $G_S$, i.e. $\Omega_I = \rho(\theta_I)$ where $\rho : \mathfrak{g}_g \rightarrow \mathfrak{g}_S$ is a Lie algebra morphism.

\subsection{Proof of consistency in the lower-dimensional supergravity}
\label{subsec:prooflowDim}

The reasoning we presented so far is only valid when the parent theory originates from a $G_S$ structure, i.e. when it admits an uplift. In this section we will extend these results dropping this condition, starting from any theory, i.e. upliftable or not. As stated earlier, for the sake of concreteness, we shall focus on the case in which the parent theory is maximal. The proof of consistency boils down to showing that the equations of motion of the parent theory, when evaluated at a $G_S$-invariant point, are themselves $G_S$-invariant. In other words, one has to show that the non-intrinsic, $G_S$ covariant pieces of the embedding tensor drop out of the equations of motion. This can be done trivially for the variation with respect to the metric (because the metric is a $G_D$ singlet).

\paragraph{Vector equations of motion}

The vector equations of motion are also $G_S$-invariant, which requires us to use the linear constraints implied by the decomposition \eqref{eq:embTensorSplit} for the vector equations of motion. Indeed, the components $T_I$ of the embedding tensor have image in $\mathfrak{g}_D \oplus \mathfrak{g}_S$ thus in the $G_S$-invariant sector the field strength simplifies to
\begin{equation}
    \mathcal{H}^M = \mathbb{P}_I{}^M \mathcal{H}^I \hspace{5mm}\text{with}\hspace{5mm} \mathcal{H}^I = dA^I +\tfrac{1}{2} \theta_{JK}{}^I A^J A^K - \frac{1}{2}\theta^{I\alpha_D}B_{\alpha_D}
\end{equation}
The matrix $\mathbb{P}_I{}^M$ is the $G_S$-invariant projectors of the vector $M$ index (transforming under $G_D$) on its $G_S$-invariant subspace, labelled by $I$. We also compute that
\begin{equation}
    D \mathcal{M}_{MN} = d \mathcal{M}_{MN} + A^I \theta_{I(M}{}^P \mathcal{M}_{N) P} = d \mathcal{M}_{MN} + A^I \theta_{I(J}{}^K \mathcal{M}_{L) K} \mathbb{P}_M{}^J \mathbb{P}_{N}{}^I.
\end{equation}
Since $\mathcal{M}_{MN}$ is $G_S$-invariant, the $\Omega_I$ contributions vanish. From there, the vector equations read
\begin{equation}
    D(\mathcal{M}_{MN} \star \mathcal{H}^N) - \frac{1}{8} T_{MN}{}^P \mathcal{M}^{NQ} \star D\mathcal{M}_{QP} = 0
\end{equation}
which reduce to
\begin{equation}
\begin{split}
    \mathbb{P}_M{}^J D(M_{IJ}\star \mathcal{H}^I) &- \frac{1}{8} \mathbb{P}_M{}^J \theta_{JN}{}^P \mathcal{M}^{NQ} \star d\mathcal{M}_{QP} \\
    &- \frac{1}{8}  \mathbb{P}_M{}^J\theta_{JN}{}^P \mathcal{M}^{NQ} \star A^I\theta_{I(P}{}^R\mathcal{M}_{Q)R} = 0
\end{split}
\end{equation}
since the only $\theta_M{}^{\alpha_D}$ component of the embedding tensor is of the form $\mathbb{P}_M{}^I \theta_I$, this equation is $G_S$-invariant .

\paragraph{Scalar equations of motion}
It is harder to show $G_S$-invariance of the scalar equations of motion due to the presence of a scalar potential. Indeed, the contribution to the equations of motion from the kinetic terms and the couplings to the scalars is easily shown to be $G_S$-invariant using the same techniques as for the vector equations of motion, using linear constraints on $T$. Conversely, showing that $\delta_s V_{|G_S\text{-inv}}$ is $G_S$-invariant is a quadratic constraint on the T-tensor and harder to show. 

Without referring to the existence of an uplift, we show here that such constraints hold for ``supersymmetric truncations" of the four-dimensional maximal supergravities with a structure group given by $G_S \subset \SU(8-\mathcal{N}) \subset \SU(8)_R$ with $\mathcal{N}\geq1$, i.e. preserving some amount of supersymmetry. This is shown by using the potential Ward identities, implied by the quadratic constraints of the parent theory, which relate the scalar potential to the quadratic expression of $A_1$ and $A_2$ as
\begin{equation}
    V \delta_A{}^B = \frac{1}{3} A^{B}{}_{CDE} A_A{}^{CDE} - 6 A_{AC} A^{BC}\,.
\end{equation}
Consider first the case where $G_S = \SU(8-\mathcal{N})$ in which case the fundamental representation of $\SU(8)$ branches as $\mathbf{8}\rightarrow (\mathcal{N},\,\mathbf{1}) \oplus (\mathbf{1},\,\mathbf{8-\mathcal{N}})$. We split the index $A \rightarrow a \,\oplus \,\alpha$ where $\alpha = 1,\,\dots,\,\mathcal{N}$ and $a = \mathcal{N}+1,\,\dots,\,8$. Then, the potential Ward identity implies that
\begin{equation}
    V \delta_\alpha^\beta = \frac{1}{3} A^{\beta}{}_{CDE} A_\alpha{}^{CDE} - 6 A_{\alpha C} A^{\beta C}\,.
\end{equation}
We can now split the index 
\begin{equation}[CDE] \rightarrow [\alpha\beta\gamma] \oplus \underbrace{[aBC]}_{\mathcal{I}}
\end{equation}
and we have that
\begin{equation}
    V \delta_\alpha^\beta = \left( \frac{1}{3} A^{\beta}{}_{\gamma\delta\epsilon} A_\alpha{}^{\gamma\delta\epsilon} - 6 A_{\alpha \gamma} A^{\beta \gamma}\right) + \left(\frac{1}{3} A^{\beta}{}_{\mathcal{I}} A_\alpha{}^{\mathcal{I}} - 6 A_{\alpha c} A^{\beta c}\right)\,.
\end{equation}
The first term is built from $G_S$-invariant quantities, while the second is not. This means that if both $A_{\alpha a}$ and $A_\alpha{}^{\mathcal{I}}$ are intrinsic, by our criterion, they must vanish or be $G_S$-invariant. We have explicitly checked that this holds in appendix \ref{App:susyCT} since the components $A_{\alpha a}$ and $A_\alpha{}^{\mathcal{I}}$ are in $W^{(\text{bad})}$, i.e. they are intrinsic and thus must vanish\footnote{The case $\mathcal{N}=1$ is a bit more subtle and requires the use of further QC on the embedding tensor for the proof of consistency as discussed in the appendices.}. This shows that
\begin{equation}
    V \delta_\alpha^\beta = \left( \frac{1}{3} A^{\beta}{}_{\gamma\delta\epsilon} A_\alpha{}^{\gamma\delta\epsilon} - 6 A_{\alpha \gamma} A^{\beta \gamma}\right) \,.
\end{equation}
In other words, we have rewritten the scalar potential using only $G_S$-invariant quantities, thus showing consistency. When $G_S$ is a proper subgroup of $\SU(8-\mathcal{N})$, the space of intrinsic torsions gets larger ($W^{int}_{G_S} \supset W^{(int)}_{\SU(8-\mathcal{N})}$) meaning that $A_{\alpha a}$ and $A_\alpha{}^\mathcal{I}$ are still intrinsic components of the torsion thus showing the result. Incidentally, when evaluated at a supersymmetric vacuum, these components of the A-matrices have to vanish \cite{Gallerati:2014xra}. This shows the lower-dimensional analogue of the Gauntlett-Varela conjecture.

We expect that similar arguments hold for other gauged supergravities as long as $G_S$ preserves some amount of supersymmetry. When there is no $G_S$-invariant gravitini we would expect that the use of the full QC (and not the subset implied by the potential Ward identity) is required to prove the consistency of such a $\mathcal{N}=0$ subsector without referring to an uplift (if it is possible at all).

\section{A worked-out example and its uplift}

To illustrate the mechanism we developed in the previous section, we consider the $D=4$ $\left[\mathrm{SO}(6)\times \mathrm{SO}(1,1)\right]\ltimes \mathbb{R}^{12}$ gauging of maximal supergravity and its $\mathcal{N}=4$ AdS$_4$ solution \cite{Gallerati:2014xra}, which admits an uplift to type IIB supergravity \cite{Inverso:2016eet}. We show that this theory does not admit a conventional truncation to pure $\mathcal{N}=4$ gauged supergravity, but instead admits an exotic one of the type described in the previous section. Since the resulting $D=4$ $\mathcal{N}=4$ gauged supergravity is realised as a subsector of a maximal $\mathcal{N}=8$ theory with a known uplift, one can directly reuse the uplift formulae of \cite{Inverso:2016eet}, restricting to modes that lie within the truncated subsector. This yields a consistent embedding of pure $D=4$ $\mathcal{N}=4$ gauged supergravity into type IIB. In this section, we present the full non-abelian uplift of the theory, extending the results of \cite{Guarino:2024gke} and matching the results of \cite{Rovere:2025jks}. As a sanity check, we computed the full scalar equations of motion in the $G_S$-invariant subsector and we show that they reduce to those of the pure $\mathcal{N}=4$ supergravity. Finally, we consider the uplift of a specific two-charge spindle solution within this theory. Since this solution contains orbifold singularities, we will comment in the next section on the regularity of its uplift. There, we will show that the ten-dimensional solution is not regularized by the uplift procedure, as is the case in \cite{Ferrero:2020twa} for example.

\subsection{The pure \texorpdfstring{$\mathcal{N}=4$}{N=4} subsector of the J-fold model}

\paragraph{Non-existence of regular invariant subsector} We compute the fermion shift matrices $A_1$ and $A_2$ at the $\mathcal{N}=4$ vacuum, and we use the $\SU(8)$ local symmetry to go to the basis in which
\begin{equation}
    A_{1\,AB}\,L={\rm diag}\left(\sqrt{2},\,\sqrt{2},\,\sqrt{2},\,\sqrt{2},\,\frac{1}{\sqrt{2}},\,\frac{1}{\sqrt{2}},\,\frac{1}{\sqrt{2}},\,\frac{1}{\sqrt{2}}\right)\,.
    \label{A10}
\end{equation}
The eigenvalues $\frac{1}{\sqrt{2}}$ correspond to the four preserved supercharges. In this basis, we can split the index  $A = 1,\,\dots,\,8$ into $a = 1,\,\dots,\,4$ and $\alpha = 5,\,\dots,\,8$. The non-vanishing components of the tensor $A_{2\,A}{}^{BCD}$ read:
\begin{equation}
A_{2\,a}{}^{bc \delta}\,L=i\,\epsilon^{abc\delta} \,\,,\,\,\,\,A_{2\,a}{}^{b \gamma\delta}\,L=-\sqrt{2}\,\delta_{ab}^{\gamma\delta}\,,\label{A20}
\end{equation}
The matrix $A_1$ is preserved by a ${{\rm O}(4)_1 \times{\rm O}(4)_2 \subset \SU(8)}$ group acting respectively on the first four and last four SU(8)-indices:
 \begin{equation}
\left(\begin{matrix}\hat{M}{}^{a}{}_{b} & {\bf 0}\cr {\bf 0} &M{}^{\alpha}{}_{\beta}\end{matrix}\right)\,,
\end{equation}
with $M,\,\hat{M} \subset {\rm O}(4)$. However, the tensor $A_2$ is only invariant under such a transformation if $M = \det(\hat{M})\, \hat{M}$. The solution is thus invariant under a ${\rm O}(4)$ group. It becomes clear that there is no subgroup of O(4) that leaves invariant the four gravitini corresponding to the preserved supersymmetry at the AdS$_4$ solution, and truncates the other four gravitini.

\paragraph{The $\SU(4)_S$ invariant subsector} The field content of pure $\mathcal{N}=4$ ungauged supergravity is obtained as an $\SU(4)$-invariant subsector of the maximal ungauged supergravity. This defines uniquely the structure group preserving exactly four gravitini and projecting out any possible vector multiplet to be the group $\SU(4)$ acting on the $a$ index. This $\SU(4)$ group is defined as a subgroup of the R-symmetry group of maximal gauged supergravity acting on the A matrices, and thus as a subgroup of the $\SU(8)_R$ group preserving $\delta_{MN}$. Since this group must preserve $\mathcal{M}_{\mathcal{N}=4} = \mathcal{V}_{\mathcal{N}=4} \cdot \mathcal{V}^T_{\mathcal{N}=4}$ the subsector we are looking for is actually the conjugated subgroup $\SU(4)_S = \mathcal{V}_{\mathcal{N}=4} \cdot \SU(4) \cdot \mathcal{V}_{\mathcal{N}=4}^{-1}$. 

For this choice of structure group, we have the branching rules
\begin{equation}
\begin{array}{ccccc}
    \Es &\rightarrow &\SU(8) &\rightarrow &\SU(4)_S \times \SU(4)_R \times \textrm{U}(1)\,,\\
    \mathbf{912} &\rightarrow &\mathbf{36} \oplus \mathbf{420} \oplus \text{ c.c.} &\rightarrow&\,\cdots\,\\
    \mathbf{56}&\rightarrow & \mathbf{28}\,\oplus\, \mathbf{28}&\rightarrow&\cdots\\
      \mathbf{133} & \rightarrow & \mathbf{63} \oplus \mathbf{70}&\rightarrow & \cdots
\end{array}
\end{equation} 
where the $\SU(8)$ representations branch as
\begin{align}
\mathbf{28} \rightarrow& (\mathbf{4},\,\mathbf{4})_0 \oplus(\mathbf{6},\,\mathbf{1})_2 \oplus(\mathbf{1},\,\mathbf{6})_{-2}\\
\mathbf{36} \rightarrow& {(\mathbf{1},\,\mathbf{10})_{-2} }\oplus (\mathbf{4},\,\mathbf{4})_0 \oplus (\mathbf{10},\,\mathbf{1})_{2}\label{36N4} \\
\mathbf{420} \rightarrow &{(\mathbf{1},\,\mathbf{\bar{10}})_{-2}}\oplus(\mathbf{4},\,\mathbf{4})_{0} \oplus(\mathbf{\bar{10}},\,\mathbf{1})_{2}\oplus (\mathbf{15},\,\mathbf{6})_{2} 
\oplus (\mathbf{6},\,\mathbf{15})_{-2}
\oplus(\mathbf{20},\,\mathbf{4})_{0} 
\oplus(\mathbf{4},\,\mathbf{20})_{0}\nonumber\\ & \oplus(\mathbf{\bar{4}},\,\mathbf{\bar{4}})_{\pm 4} \oplus(\mathbf{6},\,\mathbf{1})_{2} \oplus{(\mathbf{1},\,\mathbf{6})_{-2} }\,.\label{420N4}
\end{align}
The space $K_{SU(4)_S} = \mathbf{56}\, \otimes \,\mathfrak{su}(4)_S$ splits as
\begin{equation}
\begin{split}
\big[(\mathbf{1},\,\mathbf{6})_{-2}\,\oplus \,&(\mathbf{4},\,\mathbf{4})_0 \,\,\oplus\,(\mathbf{6},\,\mathbf{1})_2  \oplus \,\text{c. c.}\big] \otimes  (\mathbf{15},\,\mathbf{1})_0= \\
  \nonumber  &(\mathbf{15},\,\mathbf{6})_{-2} \oplus (\mathbf{36} \oplus \mathbf{20} \oplus \mathbf{4},\,\mathbf{4})_0 \oplus (\mathbf{64}\oplus \mathbf{\bar{10}} \oplus \mathbf{10} \oplus \mathbf{6},\,1)_2 \oplus \text{ c.c.}
\end{split}
\end{equation}
The only representations in \eqref{36N4}, \eqref{420N4}, and their conjugates,  which are neither ${\rm SU}(4)_S$-singlets nor in $K_{\SU(4)_S}$, will be collectively denoted by $  W^{\text{bad}}$ and are
\begin{equation}
\begin{array}{lll}
    W^{\text{bad}} &= \,(\mathbf{6},\,\mathbf{15})_{ 2} \,\oplus (\mathbf{4},\,\mathbf{20} )_0 \oplus(\mathbf{\bar{4}},\,\mathbf{\bar{4}})_{\pm 4} \oplus \text{c.c.} &\subset \mathbf{420} \oplus \text{c.c.} \,\\
    &\phantom{=}\oplus  (\mathbf{4},\,\mathbf{4})_0 \oplus \text{c.c.}&\subset \mathbf{36} \oplus \text{c.c.}
\end{array}
\end{equation}
These representations correspond to the $A$-matrices entries $A_{\alpha}{}^{\beta ab}$, $A_{a}{}^{bc\alpha}$, $A_{\alpha}{}^{abc}$, $A_{a}{}^{\alpha\beta\gamma}$ and $A_{\alpha a}$ ($\alpha$ labels the fundamental of $\SU(4)_R$ and $a$ the fundamental of $\SU(4)_S$). Following the discussion in \cite{Gallerati:2014xra}, these components of the fermion-shift matrices must all vanish at the supersymmetric vacuum. We conclude from this reasoning that the $\SU(4)_S$ subsector does describe a consistent truncation.

We insist that the consistency of the truncation relies on the specific choice of $G_S$ as embedded in $\Es$, and not only $G_S$ up to conjugation. This is why we perform the truncation with respect to the $\SU(4)_S$ subgroup and not the conjugate subgroup $\SU(4)$. 

\paragraph{Scalars and the scalar potential}

The truncated sector contains two scalars spanning the coset space $\SL(2)/\SO(2)$. This $\SL(2)$ subgroup is spanned, in $\SU(8)$ indices, by the $\SU(4)_R \times \SU(4)_S$ invariant generators 
\begin{equation}
    t^{[\alpha\beta\gamma\delta]}\hspace{5mm},\hspace{5mm}\,t^{[abcd]} = (t^{\alpha\beta\gamma\delta]})^* \hspace{5mm}\text{and}\hspace{5mm} \sum_{a=1}^4 t^{a}{}_{a}-\sum_{\alpha =5}^8 t^{\alpha}{}_\alpha\,.
\end{equation}
The coset representative of the $\SU(4)_S$-invariant scalars is thus
\begin{equation}
    \mathcal{V} = \mathcal{V}_{\mathcal{N}=4}\cdot \mathcal{V}_{\SL(2)}(\chi,\,\phi)
\end{equation}
In this truncated sector we compute that the A matrices are
\begin{equation}
A_{1\,ab}\,L=2\,f_+\,\delta_{ab}\hspace{5mm},\,\hspace{5mm}
    A_{1\,\alpha\beta}\,L=\bar{f}_+\,\delta_{\alpha\beta}\,,\label{A1pt}
\end{equation}
\begin{equation}
    \begin{split}
        &A_{2\,a}{}^{b\delta\gamma}\,L=-2\,f_+\,\delta_{ab}^{\delta\gamma}+ f_-\,\epsilon^{ab\delta\gamma }\,,\\
    A_{2\,a}{}^{bc\delta}\,L=i\,\epsilon^{abc\delta}\,,&\hspace{5mm}
    A_{2\,a}{}^{bcd}\,L= 2\,\bar{f}_-\, \epsilon^{abcd}\,,\hspace{5mm}
    A_{2\,\alpha}{}^{\beta\gamma\delta}\,L= {f}_-\, \epsilon^{\alpha \beta\gamma\delta }\,.\label{eq:AmatricesSL2}
    \end{split}
\end{equation}
where $L=g^{-1}$ is the AdS-radius and
\begin{equation}
  f_+(\chi,\phi)\equiv \frac{e^{-\frac{\phi }{2}} \left(-i \chi +e^{\phi }+1\right)}{2 \sqrt{2}}\,,\,\,\,f_-(\chi,\phi)\equiv \frac{e^{-\frac{\phi }{2}} \left(-i \chi +e^{\phi }-1\right)}{2 \sqrt{2}}\,.  
\end{equation}
Summing these contributions we end up with the following expression for the potential:
\begin{equation}
    g^{-2}\,V_{\mathcal{N}=8}(\chi,\phi)=8\,|f_-|^2-12\,|f_+|^2+3=2\,|f_-|^2-6\,|f_+|^2\,.
\end{equation}
Importantly, the $G_S$-invariant and $G_S$-covariant contributions conspire so that $V_{\mathcal{N}=8}$ coincides with the scalar potential obtained from $\mathcal{N}=4$ formulae:
$$g^{-2}\,V_{\mathcal{N}=4}(\chi,\phi)=2\,|f_-|^2-6\,|f_+|^2\,.$$
when the $\mathcal{N}=4$ A-matrices are identified with the appropriate components of the intrinsic torsion:
\begin{equation}
   g^{-1}\, A_{1\,\alpha\beta}^{\mathcal{N}=4}=\frac{3}{\sqrt{2}}\,A_{1,\,\alpha\beta}\,\,,\,\,\, g^{-1}\,A_{2\,\alpha\alpha'}^{\mathcal{N}=4}=\frac{3}{\sqrt{2}}\,A_{2\,\alpha}{}^{\beta\gamma\delta}\epsilon_{\alpha' \beta\gamma\delta}\,.
\end{equation}

This was expected following the discussion of the previous section, as the potential Ward identities imply that 
\begin{equation}
    V \delta_\alpha{}^\beta = \frac{1}{3} A^{\beta}{}_{CDE} A_\alpha{}^{CDE} - 6 A_{\alpha C} A^{\beta C}\, =( 2 |f_-|^2 - 6|f_+|^2) \delta_\alpha{}^\beta\,.
\end{equation}
This forces the non-$G_S$-invariant components to give the same kinds of contributions for the other terms proportional to $\delta_a{}^b$.

As a final check, we show, using only 4D quantities, that the $\SU(4)_S$ invariant subsector we considered is indeed a consistent truncation of the maximal gauged supergravity by considering part of the scalar equations of motion originating from the variations of the scalar potential 
\begin{equation}
   (\mathcal{E}^{\text{scal}})_s \supset \delta_s V = \mathcal{P}_s^{[ABCD]} (\mathcal{C}_{ABCD} + \epsilon_{ABCDEFGH}\mathcal{C}^{EFGH})\,
\end{equation}
where 
\begin{equation}
    \mathcal{C}_{BEFG} = A^A{}_{[BEF} A_{G]A} + \frac{3}{4} A^A{}_{D[BE}A^D{}_{FG]A}\,.
    \label{eq:BEFG}
\end{equation}
By performing explicit computations, using the A-matrices $\eqref{eq:AmatricesSL2}$, we show that only the $G_S$-invariant term contributes to this equation. In particular, the only non-zero contributions to $\mathcal{C}_{ABCD} + c.c.$ are
\begin{equation}
\begin{split}
    &\mathcal{P}_s{}^{\alpha\beta\gamma\delta} \frac{1}{2} \epsilon_{\alpha\beta\gamma\delta}\left[\chi + \frac{i}{2} \left( -e^{-\phi} + e^\phi + e^\phi \chi^2\right)\right] \\
   & \hspace{10mm}\text{and its complex conjugate}\hspace{2mm} \\
    &\mathcal{P}_s{}^{abcd} \epsilon_{abcd} \left[\chi - \frac{i}{2} \left( -e^{-\phi} + e^\phi + e^\phi \chi^2\right)\right].
\end{split}
\end{equation}
Those correspond to variations w.r.t the scalars parametrising the $\SL(2)/\SO(2)$ coset space kept in our truncation. This proves, as expected, consistency of the truncation at the level of the $D=4$ equations of motion, without requiring the existence of an uplift of the $\mathcal{N}=8$ $[\SO(1,\,1)\times \SO(6)] \ltimes \mathbb{R}^{12}$.

\subsection{Uplift of the \texorpdfstring{$\mathcal{N}=4$}{N=4} theory}
\label{subsec:uplift}

We adopt the conventions of \cite{Guarino:2022tlw} in order to describe the $\textrm{S}^1 \times \textrm{S}^5$ internal geometry in the type IIB uplift of the $\textrm{U}(1)_{\textrm{R}}^2$-invariant sector. The $\textrm{S}^{1}$ is parameterised by a periodic coordinate $\eta \in [0,\,T]$ of period $T$ whereas the $\textrm{S}^{5}$ is understood as two $2$-spheres $\textrm{S}_{i}^{2}$, with $i=1,\,2$, with polar and azimuthal angles $(\theta_i,\,\varphi_i)$ fibered over an interval $\alpha \in [0,\,\frac{\pi}{2}]$. The $\textrm{S}^{5}$ is embedded in $\mathbb{R}^6$ using embedding coordinates
\begin{equation}
\label{embedding_coordinates}
\begin{array}{lcll}
    Y_1 = \cos\alpha\,\cos\theta_1 & , & Y_4 = \sin\alpha\,\cos\theta_2 & , \\
    Y_2 = \cos\alpha\,\sin\theta_1\,\sin\varphi_1 & , & Y_5 = \sin\alpha\,\sin\theta_2\,\sin\varphi_2 & ,  \\
    Y_3 = \cos\alpha\,\sin\theta_1\,\cos\varphi_1  & , & Y_6 = \sin\alpha\,\sin\theta_2\,\cos\varphi_2 & ,
\end{array}
\end{equation}
satisfying $\sum Y_{m}^{2}=1$. The functions $Y^m$ are $\SO(3)\times \SO(3)$ equivariant functions on the sphere. We also introduce the $\SO(3)\times \SO(3)$ killing vectors on the two 2-spheres:
\begin{equation}
\begin{array}{ll}
    k_1 = \partial_{\varphi_1}\,, & k_{\hat{1}} = \partial_{\varphi_2}\,,\\
    k_2 = \cos\varphi_1\, \partial_{\theta_1} -\cot\theta_1\,\sin\varphi_1 \partial_{\varphi_1}\,,& k_{\hat{2}} =  \cos\varphi_2\, \partial_{\theta_2} -\cot\theta_2\,\sin\varphi_2 \partial_{\varphi_2}\,, \\
    k_3 = -\sin\varphi_1  \partial_{\theta_1} - \cot\theta_1 \cos \varphi_1 \partial_{\varphi_1}\,,\hspace{1cm} & k_{\hat{3}}=  -\sin\varphi_2  \partial_{\theta_2} - \cot\theta_2 \cos \varphi_2 \partial_{\varphi_2}\,,
\end{array}
\end{equation}
such that $k_m( Y^n) = \epsilon_{mnp}Y^p$. 

The Kaluza-Klein procedure imposes covariantizing the exterior derivative acting on the internal coordinates,  $y^m=(\eta,\,\alpha,\,\theta_i,\,\varphi_i)$, using
\begin{equation}
    dy^m \rightarrow Dy^m = dy^m + A^{a+} k_a(y^m) + A^{\hat{a}-} k_{\hat{a}}(y^m).
\end{equation}
this procedure relates the type IIB fields $C_p$ to untwisted fields $\bar{C}_p$ via
\begin{equation}
    \bar{C}_p \overset{{dy^m \rightarrow Dy^m}}{\longrightarrow} C_p
\end{equation}
and the untwisted fields $\bar{C}_p$ are obtained via their KK expansions coupling external forms to internal ones.

We introduce the functions $f_i$ which depend on the four-dimensional scalar $\tau$ and the $\alpha$ coordinate: 
\begin{equation}
\begin{array}{rcl}
f_0 &=& 1 + 2\, e^{-\xi}\, \cos^2\alpha \ , \\[2mm]
f_1 &=& 1 + 2\, e^\xi \,|\tau|^{2} \, \cos^2\alpha \ , \\[2mm]
f_2 &=& 1 + 2\, e^\xi \,\sin^2\alpha \ . 
\end{array}
\end{equation}
With these notations, the ten-dimensional metric is
\begin{equation}
\label{10D_metric}
\bar{ds}_{10}^2 = \Delta^{-1} \left(\tfrac{1}{2}\, ds_{4}^2 + \bar{g}_{mn} dy^m dy^n \right) \ ,
\end{equation}
where the internal metric $g_{mn}$ on $\textrm{S}^{1} \times \textrm{S}^{5}$ is given by
\begin{equation}
\label{g_internal}
\bar{g}_{mn} \, dy^{m} dy^{n} = d\eta^2  + d\alpha^2  + \dfrac{\cos^2\alpha}{f_1} ds_{\textrm{S}^2_1}^2+ \dfrac{\sin^2\alpha}{f_2} ds_{\textrm{S}^2_2}^2 \ ,
\end{equation}
The metrics on the two two-spheres $ds_{\textrm{S}^2_i}^2$ are the standard ones given by
\begin{equation}
ds_{\textrm{S}^2_i}^2  = d\theta_{i}^2 + \sin^2\theta_{i} \, {d\varphi_{i}}^2
\end{equation}
The non-singular warping factor reads
\begin{equation}
\Delta^{-4} = f_1 \, f_2 \ .
\end{equation}

The type IIB dilaton $\Phi$ and the RR axion $C_{0}$ are given by
\begin{equation}
\label{Phi&C0}
e^\Phi = \Delta^2 \, e^{-2 \eta} \, e^{\xi}\, f_0
\hspace{2mm} , \hspace{3mm} 
C_{0} = -e^{2\eta} \,\chi \, \cos(2\alpha)\,f_0{}^{-1}.
\end{equation}
To write down the two-form potentials of type IIB we introduce the rescaled 2-sphere volumes as
\begin{equation}
\label{twisted_volumes}
\begin{array}{rcll}
\widetilde{\textrm{vol}}_{1} &=& \cos^2\alpha\,f_1{}^{-1} \, \sin{\theta_1}\, d\theta_1 \wedge d\varphi_1 & , \\[2mm]
\widetilde{\textrm{vol}}_{2} &=& \sin^2\alpha\,f_2{}^{-1} \,\sin{\theta_2} \, d\theta_2 \wedge d\varphi_2 & .
\end{array}
\end{equation}

The two-form potentials $\mathbb{B}^{\alpha}=(\mathbb{B}^{1},\mathbb{B}^{2}) = (B_{2},C_{2})$ have a decomposition \`a la KK of the form
\begin{equation}
\label{B_definitions}
\bar{\mathbb{B}}^\alpha =  \,(e^{(-1)^\alpha \eta}\bar{\mathfrak{b}}^\alpha_{(0)} + \bar{\mathfrak{b}}^\alpha_{(1)} + \bar{\mathfrak{b}}^\alpha_{(2)}) \ ,
\end{equation}
involving four-dimensional scalar contributions
\begin{equation}
\label{B2&C2_scalar_contrib}
\begin{array}{rcl}
\bar{\mathfrak{b}}^1_{(0)} &=& - \sqrt{2}\,  \chi\,e^{\xi}\, \cos\alpha\,\,\widetilde{\textrm{vol}}_{1} - \sqrt{2}\, (1+e^{\xi})\,\sin\alpha\, \,\widetilde{\textrm{vol}}_{2} \ , \\[2mm]
\bar{\mathfrak{b}}^2_{(0)} &=& -\sqrt{2}\,(1+ e^\xi |\tau|^2) \, \cos\alpha\, \,\widetilde{\textrm{vol}}_{1} + \sqrt{2}\, \chi\, e^\xi\,\sin\alpha\,\widetilde{\textrm{vol}}_{2} \ ,
\end{array}
\end{equation}
as well as one-form contributions
\begin{equation}
\label{B2&C2_vector_contrib}
\begin{split}
   & \bar{\mathfrak{b}}^1_{(1)} =\frac{1}{\sqrt{2}} A^{a-} d(-e^{-\eta} Y^a)+\frac{1}{\sqrt{2}} A^{\hat{a}-} d(-e^{-\eta} Y^{\hat{a}}) \ , \\[2mm]
    &\bar{\mathfrak{b}}^2_{(1)} =\frac{1}{\sqrt{2}} A^{a+} d(-e^{\eta} Y^a)+\frac{1}{\sqrt{2}} A^{\hat{a}+} d(e^{\eta} Y^{\hat{a}}) \ ,\ ,
\end{split}
\end{equation}
and two-form contributions
\begin{equation}
\begin{split}
    \bar{b}^1_{(2)} &= \frac{1}{\sqrt{2}} (B^{ab} +\frac{1}{2} A^{a+} A^{b-})\,\epsilon_{abc}\, (-e^{-\eta} Y^c)+\frac{1}{\sqrt{2}}\,\frac{1}{2} A^{\hat{a}-} A^{\hat{b}-} \epsilon_{\hat{a}\hat{b}\hat{c}}(-e^{-\eta} Y^{\hat{c}})\\
    \bar{b}^2_{(2)} &= \frac{1}{\sqrt{2}} \frac{1}{2} A^{a+} A^{b+}\epsilon_{abc} (-e^\eta Y^c) + \frac{1}{\sqrt{2}}(B^{\hat{a}\hat{b}} +\frac{1}{2} A^{\hat{a}-} A^{\hat{b}+})\epsilon_{\hat{a}\hat{b}\hat{c}} (e^\eta Y^{\hat{c}})
\end{split}
\end{equation}
Using the technique presented in Appendix E of \cite{Rovere:2025jks}, the three-form field strengths are
\begin{equation}
    \begin{split}
        \overline{dB_2} =&  \mathcal{D} \bar{B}_2\\
        =& -\sqrt{2}\, d\left[e^{-\eta} \chi e^\xi \cos \alpha\, \tilde{\text{vol}}_1 + (1+e^\xi)\sin\alpha \,\tilde{\text{vol}}_2 \right]\\
        &+\mathcal{H}^{\hat{a}-} \left(\frac{1}{\sqrt{2}} d(-e^\eta Y^{\hat{a}} )+ (1+e^\xi) \sin\alpha\, \iota_{k_{\hat{a}}}\tilde{\text{vol}_2} \right)\\
        &+\mathcal{H}^{a-}\left(\frac{1}{\sqrt{2}} d(-e^\eta Y^a)\right)\\
        &+ \underbrace{\mathcal{H}^{ab} \epsilon_{abc}}_{=0\text{ on-shell}} \left(\cdots\right)\,.
    \end{split}
\end{equation}
\begin{equation}
    \begin{split}
        \overline{dC_2} =&  \mathcal{D} \bar{C}_2\\
        =& -\sqrt{2}\, d\left[e^{\eta}(1+e^\xi|\tau|^2) \cos \alpha\, \tilde{\text{vol}}_1 -\chi e^\xi \sin\alpha \,\tilde{\text{vol}}_2 \right]\\
        &+\mathcal{H}^{a+}\left(\frac{1}{\sqrt{2}} d(e^\eta Y^a)+ (1+e^\xi|\tau|^2)\cos\alpha\, \iota_{k_a}\tilde{\text{vol}}_1\right)\\
        &+\mathcal{H}^{\hat{a}+} \left(\frac{1}{\sqrt{2}} d(e^\eta Y^{\hat{a}} ) \right)\\
        &+ \underbrace{\mathcal{H}^{ab} \epsilon_{abc}}_{=0\text{ on-shell}} \left(\cdots\right)\,.
    \end{split}
\end{equation}

Finally, the self-dual five-form field strength reads
\begin{equation}
\begin{array}{lll}
\bar{\widetilde{F}}_5 &=&-(1+\bar{\star})\Big[- \Big((2 + V) \sin(2\alpha) d\eta \\
&&  + \big( 2 (2 + e^\xi) \cos 2\alpha + V \cos^2\alpha \big) d\alpha \Big)\wedge \tilde{\text{vol}}_1\wedge \tilde{\text{vol}}_2\\
&&+\sin(2\alpha) (d\xi - e^{2\xi}\chi d\chi)  \wedge \tilde{\text{vol}}_1\wedge \tilde{\text{vol}}_2 \\[2mm]
&&+\mathcal{H}^{\hat{a}-} Y^{\hat{a}}(\cos\alpha d\eta + \sin\alpha d\alpha)\wedge \tilde{\text{vol}}_1 \\
&&+\mathcal{H}^{\hat{a}+}  \left(Y^{\hat{a}}(\cos\alpha d\eta - \sin\alpha d\alpha) \tilde{\text{vol}}_2 + \epsilon_{\hat{a}\hat{b}\hat{c}} Y^{\hat{b}} d(Y^{\hat{c}}/\sin\alpha) \cos\alpha \,d\eta \wedge d\alpha\right)\\[2mm]
&&+\mathcal{H}^{a+} Y^a (\sin\alpha d\eta + \cos\alpha d\alpha)\wedge \tilde{\text{vol}}_2\\
&&+\mathcal{H}^{a-} \left(Y^{a}(-\sin\alpha d\eta +\cos\alpha d\alpha) \tilde{\text{vol}}_1 + \epsilon_{abc} Y^{b} d(Y^c/\cos\alpha) \sin\alpha \,d\eta \wedge d\alpha\right)\Big]\,.
\end{array}
\label{tildeF_5}
\end{equation}

\paragraph{Reduction of the equations of motion}
With this ansatz, one can check explicitly that the type IIB equations of motion do reduce to those of the four-dimensional pure $\mathcal{N}=4$ theory as presented in section \ref{sec:D=4N=4}. This computation is not particularly enlightening, and the methods used to perform it are spelt out in Appendix E of \cite{Rovere:2025jks}. As an example, we show here that the Bianchi identify for $F_5$ reduces to the four-dimesional scalar equations of motion. We consider
\begin{equation}
    d \tilde{F}_5 = dB_2 \wedge dC_2\,.
\end{equation}
We can factorise out the KK twist to obtain 
\begin{equation}
    \mathcal{D} \bar{\tilde{F}}_5 = \mathcal{D}\bar{B_2} \wedge \mathcal{D}\bar{C}_2\,.
\end{equation}
We can then factorise this equation according to its rank in the external space. Since the ExFT dictionary only required to add external three- and four-forms by hand, in order to satisfy the twisted-self duality equations, we will solve the Bianchi identities off-shell for the zero-form, one-form and two-form components of these equations. For the three-form contribution, only the three-form component of 
\begin{equation}
    d[\star_4 (d\xi - e^{2\xi}\chi d\chi)\,\,\bar{\star}_6( \sin(2\alpha) \tilde{vol}_1 \tilde{vol}_2)]_{|\text{ext rk}=3}
\end{equation}
could be non-trivial. This reads 
\begin{equation}
    \star_4 (d\xi - e^{2\xi}\chi d\chi)\,\,d(d\alpha \wedge d\eta) = 0
\end{equation}
which is satisfied off-shell. Finally, the external four-form components of the five-form Bianchi identity reads, upon using the twisted self-duality relation for $\mathcal{H}$, as 
\begin{equation}
    \text{vol}_{ext} 8 \cos\alpha\,\sin\alpha \left[\left(\text{e.o.m. of }\xi\right) + \chi (\text{e.o.m. of }\chi)\right] = 0\,.
\end{equation}
which is satisfied for solutions of the $\mathcal{N}=4$ theory. Equivalent methods can be used to compute the equations of motion for the other type IIB fields.

\subsection{Uplift of the two charges near-horizon spindle}

We will now consider the solutions of the $D=4$ $\mathcal{N}=4$ $\SO(4)$-gauged supergravity presented in \cite{Ferrero:2021ovq}. These solutions are conjectured to capture the near-horizon limit of rotating and accelerating charged black holes with AdS$_4$ asymptotics. Here, we will summarise the construction of the non-rotating purely magnetic solutions, and we will comment on the regularity of their uplifts. 

\subsubsection{The solution and quantisation conditions}
The solution of \cite{Ferrero:2021ovq}, in the limit of vanishing rotation with only magnetic charges turned on reduces to a geometry of the form $\mathrm{AdS}_2 \times \spindle$, where $\spindle$ is a topological sphere with two orbifold singularities. In adapted coordinates $(t,\,\rho,\,w,\,z)$ this solution reads
\begin{align}
\mathrm{d}s^2_4
  &= \frac{1}{4}\,\Lambda(w)\!\left(-\rho^{2}\,\mathrm{d}t^{2}
      + \frac{\mathrm{d}\rho^{2}}{\rho^{2}}\right)
     \;+\; \frac{\Lambda(w)}{Q(w)}\,\mathrm{d}w^{2}
     \;+\; \frac{Q(w)}{4\,\Lambda(w)}\,\mathrm{d}z^{2},
\\[4pt]
A_i &= \frac{w}{w+q_i}\,\mathrm{d}z,
\qquad
e^{\xi} = \frac{w+q_{1}}{\,w+q_{2}\,},
\qquad
\chi = 0,
\label{eq:solSpindle}
\end{align}
where
\begin{align}
\Lambda(w)=(w+q_{1})(w+q_{2})\hspace{5mm}\text{and}\hspace{5mm}
Q(w)=\Lambda(w)^{2}-4w^{2}.
\end{align}

The uplift is obtained by identifying in the formulae given in Sect. \ref{subsec:uplift}, the metric and scalar fields as in \eqref{eq:solSpindle}. The vectors are identified as
\begin{equation}
    \begin{split}
        &A^{1+} = \frac{w}{w+q_1} \mathrm{d}z \hspace{1cm} A^{1-}= - \frac{q_1}{2} \rho \mathrm{d}t\\
        &A^{\hat{1}-}= \frac{w}{w+q_2} \mathrm{d}z\hspace{1cm}   A^{\hat{1}+} =  - \frac{q_2}{2} \rho \mathrm{d}t
    \end{split}
\end{equation}
The magnetic duals to the vectors are obtained by using the twisted self-duality equation \eqref{eq:selfdualityeqN4} and setting the four-dimensional two-forms $B^{ij}$ to zero. All other vector fields are set to zero.

Depending on the values of $q_1$ and $q_2$, the ranges of coordinates where the metric is regular varies. For $\spindle$ to describe a spindle we must impose that $2> q_1,\,q_2 >0$ and $\Delta_- >0$ where we introduced the shorthands $\Delta_\pm = 4 \pm 4(q_1 + q_2) + (q_1 - q_2)^2$. These conditions imply that $Q$ has the following, ordered, four real roots
\begin{equation}
    \begin{split}
        &w_1 = \frac{1}{2}\left(-2 -(q_1 + q_2) - \sqrt{4+4(q_1+q_2) + (q_1-q_2)^2}\right),\,\\
        &w_2 = \frac{1}{2}\left(-2 -(q_1 + q_2) + \sqrt{4+4(q_1+q_2) + (q_1-q_2)^2}\right),\,\\
        &w_3 = \frac{1}{2}\left(2 -(q_1 + q_2) - \sqrt{4-4(q_1+q_2) + (q_1-q_2)^2}\right),\,\\
        &w_4 = \frac{1}{2}\left(2 -(q_1 + q_2) + \sqrt{4-4(q_1+q_2) + (q_1-q_2)^2}\right).\,
    \end{split}
\end{equation}
and the metric is positive definite when $w \in [w_2,\,w_3]$.

We must now introduce quantization conditions on the period of $z$ as well as on the parameters $q_1$ and $q_2$ in order to obtain proper orbifold singularities at the poles of the spindle. Near the poles, the metric reads
\begin{equation}
    ds_{\includegraphics[height=6pt]{mySpindle.png}}^2 \approx dr^2 + \frac{1}{4} \Delta_\pm \,r^2 dz^2 + O(r)^3
\end{equation}
where the $\Delta_+$ is the appropriate factor at $w_2$, and $\Delta_-$ the one at $w_3$. We must thus impose
\begin{equation}
    \frac{\sqrt{\Delta_\pm}}{2} \Delta z = \frac{2\pi}{n_\pm}\,.
\end{equation}
for $n_\pm \in \mathbb{Z}_{>0}$.
Finally, the vector fields $A_i$ must be connections on regular principal bundles over the orbifold. This imposes the following quantisation condition:
\begin{equation}
    \frac{1}{2\pi} \int_\Sigma F_i = \frac{\Delta z}{2\pi}\, \left[\frac{w}{w+q_i}\right]_{w_-}^{w_+}=\frac{p_i}{n_+n_-}
\end{equation}
for $p_i$ an integer. We evaluate this integral and obtain
\begin{equation}
\begin{split}
      &\frac{1}{2\pi}\int_\Sigma F_1 = \frac{1}{2}\left(\frac{1}{n_+}-\frac{1}{n_-} + \sqrt{\frac{2}{n_-^2}+\frac{2}{n_+^2} -\frac{\Delta z^2}{\pi^2}} \right)\\
      &\frac{1}{2\pi}\int_\Sigma F_2= \frac{1}{2}\left(\frac{1}{n_+}-\frac{1}{n_-} - \sqrt{\frac{2}{n_-^2}+\frac{2}{n_+^2} -\frac{\Delta z^2}{\pi^2}} \right)\\
\end{split}      
\end{equation}
From this, we read that
\begin{equation}
\label{eq:pQuant}
    p_i = \frac{1}{2}\left(n_- -n_+ \mp \underbrace{\sqrt{2 n_+^2+2 n_-^2 - \left(\frac{n_+n_-\Delta z}{2\pi}\right)^2}}_{=:k}\right)\,.
\end{equation}
and we obtain
\begin{equation}
    \begin{split}
    \label{eq:quantFinal}
        &q_1 = \frac{n_-^2 - n_+^2 + k \sqrt{2(n_+^2+n_-^2)-k^2}}{2(n_+^2+n_-^2)-k^2} \\
        &q_2= \frac{n_-^2 - n_+^2 - k \sqrt{2(n_+^2+n_-^2)-k^2}}{2(n_+^2+n_-^2)-k^2} \\
        &\Delta z = \pi n_-^2 n_+^2\sqrt{2(n_+^2+n_-^2)-k^2}
    \end{split}
\end{equation}
which leads to regular solutions whenever $n_+$, $n_-$ are coprime integers and $k$ is an integer such that the $p_i$ are themselves co-prime and coprime with $n_\pm$. We stress that these regularity conditions are conditions for the ``orbifold" regularity at the level of the four-dimensional fields. In the next subsection, we will recall why the precise quantisation condition we imposed must hold and how this implies (or not) the regularity of the uplift.

\section{Regularity of uplifts of orbifold solutions}
\label{sec:Regularity}

Having reviewed an orbifold solution in four-dimension, one should wonder if its singularities are resolved by the uplift procedure. We state the following criterion for the regularity of the uplift:

\emph{The uplift of a lower-dimensional solution on an orbifold $\mathcal{O}$ with a compact gauge group $G$ is smooth if and only if
\begin{enumerate}
    \item The gauge connection on the orbifold is a connection on a smooth $G$-principal {(orbi-)bundle} on $\mathcal{O}$.
    \item The isotropy groups of the orbifold, $\Gamma$, which naturally embed in $G$, act freely on $\Mint$.
\end{enumerate}}
\noindent The first condition imposes quantisation conditions on the various charges of the solution, while the second condition is a constraint on the internal manifold, which admits a $G$ action as was shown in \cite{Rovere:2025jks}. We will motivate this result here, and we will apply it to three examples:
\begin{itemize}
    \item The uplift of single-charge \cite{Ferrero:2020twa,Cassani:2021dwa} and multicharges \cite{Ferrero:2021ovq,Ferrero:2021etw,Couzens:2021rlk} $D=4$ non-rotating spindles in M-theory on $SE_7$ and $S^7$, respectively.
    \item The uplift of the two-charges $D=7$ spindle to M-theory on $S^4$ \cite{Ferrero:2021wvk,Bomans:2024mrf}.
    \item The uplift of the two-charges spindle in type IIB we built here.
\end{itemize}
We will recover various results in the literature for the first two cases, but also show why the type IIB uplift is not regular. Indeed, upon enforcing the appropriate quantisation conditions on the spindle charges, we will show that the ten-dimensional solution includes orbifold defects localised on eight disconnected four-dimensional subspaces of the full ten-dimensional geometry, and we will characterise the orbifold singularities near these defects.

\subsection{The regularity condition}

Let us first consider orbifold solutions in the low-dimensional gauged supergravity. In this context, the low-dimensional vector fields should be understood as a connection on a $G$-principal bundle $P$ over an orbifold $\mathcal{O}$. We consider the case where, upon imposing the appropriate quantisation conditions on the various parameters of the solution, $P$ is a manifold, and the action of $G$ on $P$ is locally free. This construction endows $P$ with an orbibundle structure whose projector is denoted $\pi: P \rightarrow \mathcal{O}=P/G$. The stabiliser of a point $p \in P$ is a discrete group $\Gamma_p = \text{Stab}(p)$ isomorphic to $\Gamma_{o=\pi(p)}$, the isotropy group of $o\in\mathcal{O}$. Up to conjugation, this embeds $\Gamma_o \in G$. Finally, we note that the local trivializations of $P$ are of the form
\begin{equation}
    U_{o=\pi(p)} \times_{\Gamma_p} G\,.
\end{equation}
where $U_o$ is an open subset of $\mathbb{R}^n$ admitting an action of $\Gamma_p$.

We can now study the total manifold of the uplifted solution. This total manifold is the associated bundle
\begin{equation}
    \Mtot = P\times_{G} \Mint
\end{equation}
where $\Mint$ must admit an effective, but not necessarily free, $G$-action. Points in $\Mtot$ are given by equivalence classes of pairs $(p,\,m) \in P \times \Mint$ up to a $G$ action given by $[p,\,m] \cong [p\cdot g^{-1},\,g\cdot m]$  $ g\in G$). In particular, the quotient will be smooth if the stabilisers of all such pairs are trivial i.e.
$$G_{(p,\,m)}=\{e\}\,.$$
We must thus evaluate whether $G_{p}\,\cap\,G_{m} = \Gamma_p \,\cap\,G_{m}$ is trivial for any pair $(p, \,m)$. When the action of $G$ is free on $\Mint$, or when $P$ is a regular principal bundle (i.e. $G$ acts freely on $P$), the uplift is always regular.

Studying local models for $P$ and $M_{tot}$, this discussion can be made more intuitive. Near a point $p$ with $\pi(p) = o$ and with stabiliser $\Gamma_p \in G$, we can build a local model $P_{|\pi^{-1}(U_o)} \cong U_o \times_{\Gamma_p} G$. Then the uplift of this patch will locally look like
\begin{equation}
     U_o \times_{\Gamma_p} \times G \times_{G} M_{int} \cong U_o \times_{\Gamma_p} M_{int}\,.
\end{equation}
It becomes clear that this patch is smooth if and only if $\Gamma_p$ acts freely on $M_{int}$ as we claimed.

To summarize, in order to prove the regularity of the uplift, one must first build a smooth $G$-principal bundle $P$ over the orbifold $\mathcal{O}$. Then, one must check that the gauge connection is a well-defined connection for this principal bundle. For $\Urm(1)$ gauging, this simply requires that the Chern class of the connection matches that of a connection on $P$, which is quantised, see eq.\eqref{eq:quantFinal} for an example. The second step is to check whether the embedding of $\Gamma_o$ in $G$ acts freely on $\Mint$, showing or not the regularity of the uplift.

\subsection{Examples}

As a warm-up, we will first consider how our criterion applies to the uplift of the single charge spindle on $SE_7$ manifolds, regular or not. As shown in \cite{Ferrero:2020twa}, the $\Urm(1)$-gauge connection for such four-dimensional spindles originates from smooth Lens spaces $L(p,\,1)$. Moving on to multi-charge spindles, we show that the full $\Urm(1)^r$ connection can be built from the diagonal pullback of each of the individual $\Urm(1)_I$ bundle $L(p_I,\,1)$ corresponding to each charge and is still smooth. This forces the isotropy group of the orbifold to embed \emph{diagonally} in $\Urm(1)^r$ and allows us to show that the uplift is once again smooth. Conversely, we also recover the result of \cite{Ferrero:2021wvk,Bomans:2024mrf}, that certain seven-dimensional spindle solutions uplift to manifolds with $\mathbb{C}^3/\mathbb{Z}_{n_\pm}$ orbifold singularities. This illustrates how our criterion applies for singular uplifts. Finally, we will show that the type IIB uplifts of the two-charges spindles must contain eight $\mathbb{C}^3/\mathbb{Z}_{n_\pm}$ singularities. 

\paragraph{Single charge spindles} 
We consider the solution of \cite{Ferrero:2020twa} with $j=0$ and $a=q/2$. This solution can be obtained from \eqref{eq:solSpindle} by identifying $q_1 = q_2 = q$, truncating further the $\mathcal{N}=4$ theory to its pure $\mathcal{N}=2$ subsector, removing the constant scalars and identifying the two $\mathcal{N}=4$ $\mathrm{U}(1)$ vector fields. As explained in the appendix A of \cite{Ferrero:2020twa}, the $\mathrm{U}(1)$ bundle over the  spindle can be modelled as $S^3/\mathbb{Z}_p$ with $gcd(p,\,n_\pm)=1$ and $gcd(n_+,\,n_-)=1$. The $\mathrm{U}(1)$ (and its $\mathbb{Z}_p$ subgroup) action on $S^3/\mathbb{Z}_p$ is defined via its weighted action on $\mathbb{C}^2$:
\begin{equation}
    \lambda\cdot(z_1,\,z_2) = (\lambda^{n_+} z_1,\,\lambda^{n_-} z_2)\hspace{5mm}\text{for}\hspace{5mm}(z_1,\,z_2) \in S^3 \subset \mathbb{C}^2\text{ and }\lambda \in U(1)\,.
    \label{eq:LensSpaceU1Action}
\end{equation}
Since the base space is an orbifold, it comes with no surprise that the stabilisers of the north ($z_1=0$) and south ($z_2= 0$) poles of the spindle are non-trivial:
\begin{equation}
   \mathrm{U}(1)_{z_1=0} = \mathbb{Z}_{n_-} \hspace{5mm}\text{and}\hspace{5mm}\mathrm{U}(1)_{z_2=0} = \mathbb{Z}_{n_+}\,.
\end{equation}
 We can thus identify the base space with the orbifold $\spindle=\mathbb{WCP}_{[n_+,\,n_-]}$. Then taking into account all global issues, the U(1) connection of the solution lifts to a connection for this total space with the appropriate Chern number 
 \begin{equation}
     \int_\Sigma F = p/(n_+ n_-)\,.
 \end{equation}

The slice representation theorem \cite{Meinrenken2003} provides two local models near the poles. At $z_1 = 0$ the local model is 
\begin{equation}
    \mathbb{C} \times_{\mathbb{Z}_{n_-}}(U(1)/\mathbb{Z}_p)\,.
\end{equation}
The action of $\omega_{n_-} = e^{2 \pi i/n_-}$, a generator for the $\mathbb{Z}_{n_-}$ group, in this local model is inherited from \eqref{eq:LensSpaceU1Action} and reads
\begin{equation}
    \omega_{n_-} \cdot (z,\,e^{i\theta}) = (\omega_{n_-}^{n_+} z,\, \omega_{n_-}^{p}e^{i\theta})\,.
\end{equation}
The factor $p$ in the second term takes care of the periodicity of $\theta \in [0,\,2\pi[$ parametrising $\Urm(1)/\mathbb{Z}_{p}$. It is possible to use another generator of $\omega_{n_-}^{m_-}$ of $\mathbb{Z}_{n_-}$ such that the action on the first factor is more natural. Choosing
\begin{equation}
    m_- = n_+^{-1}\,\, \text{mod} \,n_-\hspace{5mm}\Leftrightarrow\hspace{5mm} m_- n_+ - m_+ n_- = 1 \hspace{5mm}\text{(Bézout-Euclid)}
\end{equation}
The action reduces to
\begin{equation}
    \omega^{m_-}_{n_-}(z,\,\theta) = (\omega_{n_-} z, \omega_{n_-}^{p\,m_-}e^{i\theta})\,,
    \label{eq:chargesSpindles}
\end{equation}
And so the fibre has charge $p\,m_{-}$. At the other pole, the roles of $(n_+,\,m_+)$ and $(n_-,\,m_-)$ are exchanged.

The regularity of the total space can then be inferred from the regularity of the $\mathrm{U}(1)$ action on $M_{\text{int}}$ in the $11$D uplift. When uplifting the solution as a compactification of M-theory on a \emph{regular} Sasaki-Einstein manifold, the full $\Urm(1)$ action is free and thus so are the $\mathbb{Z}_{n_+}$ and $\mathbb{Z}_{n_-}$ actions. As such, the uplift is smooth. When the SE$_7$ is quasi-regular, it admits a locally free U(1) action, i.e. there exists singular points, fixed under a $\mathbb{Z}_q \subset \Urm(1)$ action. As such the $\Urm(1)$-action on $M_{\text{ext}} \times M_{\text{int}}$ is free whenever $\text{gcd}(n_\pm,\,q) = 1$ and in such case the uplift will be smooth. Conversely, when $\text{gcd}(n_\pm,\,q) = n >1$, the uplift will contain a $\mathbb{Z}_n$ orbifold singularity.

\paragraph{Multi-charge spindles}
We can now move to the multicharge spindle solution \cite{Ferrero:2021ovq}, which we reviewed in detail in the previous subsection. For the  $\Urm(1)^2$ gauge theory, we must build a regular $\Urm(1)^2$ principal bundle over $\Sigma$. The quantisation conditions for the various charges are given by equations \eqref{eq:pQuant} and \eqref{eq:quantFinal}. Using the results concerning single-charge spindles, with each of the $\Urm(1)$ gauge fields, we can associate a principal $\Urm(1)$ orbibundle $P_i = S^3/\mathbb{Z}_{p_i}$ over the orbifold. The total principal bundle, from which the full $\Urm(1)^2$ connection originate is then
\begin{equation}
    P_1 \times_O P_2 = \left\{(p_1,\,p_2)| \pi(p_1) = \pi(p_2)\right\}\,.
\end{equation}
In the appendix \ref{app:orbifolds}, we show that this is a smooth space and that the orbifold isotropy groups $\mathbb{Z}_{n_+}$ and $\mathbb{Z}_{n_-}$ embed \emph{diagonally} in $\Urm(1) \times \Urm(1)$. Using our regularity criterion, the uplift of this solution on $S^7 \subset \mathbb{C}^2 \times \mathbb{C}^2$ admits a free action for both $\mathbb{Z}_{n_\pm}$ isotropy groups. 

The same construction, and result, generalises to the four-charge spindles of \cite{Couzens:2021rlk}, obtained from the abelian subsector of the seven-sphere reduction, this time embedding $\mathbb{Z}_{n_\pm}$ diagonally on $\Urm(1)^4$ acting on each of $\mathbb{C}$ factors of $\mathbb{C}^4 \supset S^7$ with weight $p_I$.

\paragraph{The class-$\mathcal{S}$ spindle} Consider the uplift spindle solution of $SO(5)$-gauged D=7 supergravity of \cite{Ferrero:2021wvk} whose internal space is $S^4$. This solution is obtained in a $\Urm(1)^2 \subset \SO(5)$ invariant subsector of this supergravity, retaining two abelian vector fields. The seven-dimensional solution is of the form \begin{equation}
    M_7 = AdS_5 \times \spindle
\end{equation} where $\spindle$ is, once again, a spindle with two singularities of order $n_+$ and $n_-$. 

As for the four-dimensional two-charge black hole, upon enforcing the appropriate quantisation conditions, this manifold can be seen as the base space of a $\Urm(1)^2$ orbibundle characterised by two integers $p_{1, \,2}$. The isotropy groups $\mathbb{Z}_{n_{\pm}}$ are embedded diagonally in $\Urm(1)^2 \subset \SO(5)$. The internal four sphere can be considered as the set of points $(z_1,\,z_2,\,x) \in \mathbb{C}^2 \times \mathbb{R}$ with $|z_1|^2 + |z_2|^2 + x^2 = 1$. The diagonal $\Urm(1)$ group has a trivial action on $S^4$ whenever $x^2=1$. This implies that the uplift of this spindle solution will contain four five-dimensional defects, spanning the AdS$_5$ direction. Near these orbifold singularities, orbifold factor is of the form $\mathbb{C}^3/\mathbb{Z}_{n_{\pm}}$ where the action of $\mathbb{Z}_{n_\pm}$, following \eqref{eq:chargesSpindles}, has charges $(1,\,\pm p_1 m_\pm,\, \pm p_2 m_\pm)$. The signs depending on whether $x=1$ or $x=-1$. As we should, we recover the results of \cite{Ferrero:2021wvk,Bomans:2024mrf}.

\paragraph{Type IIB uplift}
Without having to perform the explicit uplift and studying singularities in a coordinate system, our criterion states that the type IIB uplift of the solution \eqref{eq:solSpindle} contains 8 co-dimension 6 singularities. These singularities are localised at the poles of the two two-spheres and at the tips of the spindle. Indeed, the diagonal $\Urm(1)$ action on $\Mint$ is free except at the poles of both spheres, where the $\Urm(1)^2$ action is trivial. This implies that the $\mathbb{Z}_{n_\pm}$ action is trivial, hence the singularities. These singularities span the AdS$_2$ factor and the annulus parametrised by $(\alpha,\,\eta)$. Near the singularities, these conical defects can be represented as $(\mathbb{C}\times \mathbb{C}_1\times\mathbb{C}_2)/\mathbb{Z}_{n_\pm}$ where each $\mathbb{C}$ factor represents a local coordinate near the pole of $\spindle$, $S^2_1$ and $S_2^2$ respectively. The $\mathbb{Z}_{n_\pm}$ isotropy groups act (anti-)diagonally on each $\mathbb{C}$ factor (depending whether we sit at the north or south pole of a sphere) as \eqref{eq:chargesSpindles}:
\begin{equation}
    \omega_{n_\pm}\cdot (z,\,z_1,\,z_2) = (\omega_{n_\pm} z,\,\omega_{n_\pm}^{\pm p_1 m_\pm}z_1,\,\omega_{n_\pm}^{\pm p_2 m_{\pm}}z_2)\,.
\end{equation}

\section{Scanning for other consistent truncations}
In this section, we apply the general method discussed in Section \ref{sec:3} to search for other consistent truncations of the J-fold supergravity, which include the pure  $\mathcal{N}=3$ supergravity and certain $\mathcal{N}=2$ models with vector multiplets only. As for the latter class of models, we limit our analysis to a restricted number of possibilities. We anticipate here our result: the consistent truncations with less than sixteen supercharges, within the considered models, are:
\begin{itemize}
    \item Pure  $\mathcal{N}=3$ supergravity for $\delta=1,\,\chi=0$ (i.e. truncation of the pure $\mathcal{N}=4$ truncation discussed earlier), discussed in  Appendix \ref{App:susyCT};
     \item Pure  $\mathcal{N}=2$ supergravity for generic $\delta,\,\chi$, discussed in  Appendix \ref{App:susyCT};
     \item A specific $\mathcal{N}=2$ supergravity coupled to one vector multiplet (the $t$-model) for generic $\chi$ and $\delta=1$.
\end{itemize}
It is useful to choose the basis of the fundamental representation of ${\rm SU}(8)$ so that the matrix $A_{1\,ij}$ has the following form:
{\scriptsize \begin{equation}
    A_{1\,ij}=\left(
\begin{array}{cccccccc}
 -1-i & 0 & 0 & 0 & 0 & 0 & 0 & 0 \\
 0 & -1-i & 0 & 0 & 0 & 0 & 0 & 0 \\
 0 & 0 & \frac{i \left(\delta ^2+\delta +2\right)}{2 (\delta -i)} & \frac{i \chi }{\sqrt{2} \sqrt{\delta ^2+1}} & 0 & 0 & 0 & 0 \\
 0 & 0 & \frac{i \chi }{\sqrt{2} \sqrt{\delta ^2+1}} & -\frac{i ((\delta -1) \delta +2)}{2 (\delta +i)} & 0 & 0 & 0 & 0 \\
 0 & 0 & 0 & 0 & \frac{i \left(\delta ^2+\delta +2\right)}{2 (\delta -i)} & \frac{i \chi }{\sqrt{2} \sqrt{\delta ^2+1}} & 0 & 0 \\
 0 & 0 & 0 & 0 & \frac{i \chi }{\sqrt{2} \sqrt{\delta ^2+1}} & -\frac{i ((\delta -1) \delta +2)}{2 (\delta +i)} & 0 & 0 \\
 0 & 0 & 0 & 0 & 0 & 0 & -\frac{1}{2}+\frac{i}{2} & 0 \\
 0 & 0 & 0 & 0 & 0 & 0 & 0 & -\frac{1}{2}+\frac{i}{2} \\
\end{array}
\right)\,.
\end{equation}}
At the $\mathcal{N}=4$ vacuum ($\chi=0$ and $\delta=1$), and the $\mathcal{N}=2\&{\rm U}(2)$ one, the latter matrix read
\begin{align}
   \mathcal{N}=4&:\,\,\, A_1\bar{A}_1={\rm diag}\left(2,2,2,\frac{1}{2},2,\frac{1}{2},\frac{1}{2},\frac{1}{2}\right)\,,\\
    \mathcal{N}=2\&{\rm U}(2)&:\,\,\, A_1\bar{A}_1={\rm diag}\left(2,2,1,1,1,1,\frac{1}{2},\frac{1}{2}\right)\,,\\
\end{align}
In the former case, the preserved supersymmetries correspond to the entries $4,6,7,8$ and the broken ones to $1,2,3,5$. In the latter case the preserved supersymmetries correspond to the entries $7,8$.
As far as the pure $\mathcal{N}=3$ and $\mathcal{N}=2$ truncations are concerned, their consistency within the maximal theory is proven in Appendix \ref{App:susyCT}.
\subsection{The \texorpdfstring{$t^3$}{t\^ 3}-model}
The structure group in this case is $G_S={\rm USp}(6)$, subgroup of the ${\rm SU}(6)$ group defining the pure $\mathcal{N}=2$ truncation. It describes $\mathcal{N}=2$ supergravity coupled to one vector multiplet, whose complex scalar field spans the scalar manifold: $$\mathcal{M}_{\rm scal.}=\frac{{\rm SL}(2,\mathbb{R})}{{\rm SO}(2)}\,.$$
This model is part of a consistent truncation of the maximal ungauged  theory, in which its scalar manifold is completed by a quaternionic-K\"ahler manifold of quaternionic dimension 7, as follows:
$$\frac{{\rm SL}(2,\mathbb{R})}{{\rm SO}(2)}\times\frac{{\rm F}_{4(4)}}{{\rm USp}(6)\times {\rm SU}(2)}\subset \frac{{\rm E}_{7(7)}}{{\rm SU}(8)}\,.$$

The branching with respect to ${\rm USp}(6)$ are readily obtained from those relative to ${\rm SU}(6)$ by decomposing the relevant 
${\rm SU}(6)$-representations. I particular, the fundamental representation of ${\rm SU}(8)$, under ${\rm USp}(6)_S\times {\rm U}(2)_R$, branches as follows
\begin{align}
    \mathbf{8} \rightarrow (\mathbf{6},\,\mathbf{1})_{-1} \oplus (\mathbf{1},\, \mathbf{2})_{+3}\,.
\end{align}
Applying the general procedure described earlier, we find that the obstruction for this model to be a consistent truncation of the J-fold theory is defined by the following representations:
$$W^{(\text{bad})}={\bf ({14}',2)}_{+2}\,\left[A_a{}^{bc\alpha}\right]\oplus {\bf ({14},1)}_{+6}\,\left[A_a{}^{b\alpha\beta}\right]\oplus c.c\,.$$
Inspection of the components of $A_1,\,A_2$ tensors, we could not find a consistent embedding of ${\rm USp}(6)$ in ${\rm SU}(6)$ for which the above representations vanish. 
\subsection{The \texorpdfstring{$t$}{t}-model truncation}
This model describes $\mathcal{N}=2$ supergravity coupled to one vector multiplet, whose complex scalar spans the scalar manifold:
$$\mathcal{M}_{{\rm scal}}=\frac{{\rm SU}(1,1)}{{\rm U}(1)}\,.$$
It is part of a consistent truncation of the maximal ungauged theory within which the above manifold is completed by a dimension-6 quaternionic-K\"ahler manifold, as follows:
\begin{equation}
\frac{{\rm U}(1,1)}{{\rm U}(1)^2}\times\frac{{\rm SO}(4,6)}{{\rm SO}(4)\times {\rm SO}(6)}\,,
\end{equation}
From the isotropy group of the above quaternionic manifold, we can read off the structure group defining the $t$-model truncation:
 $${\rm SO}(4)\sim {\rm SU}(2)_R\times {\rm SU}(2)_S\,\,,\,\,\,{\rm SO}(6)\sim {\rm SU}(4)_S\,.$$
The structure group is thus:
\begin{equation}
    G_S={\rm SU}(2)_S\times {\rm SU}(4)_S\,,
\end{equation}
With respect to the group $G_S\times {\rm SU}(2)_R\times  {\rm U}(1)^2$ the fundamental representation of ${\rm SU}(8)$ branches as follows:
\begin{equation}
\label{branching wrt SU(6)*SU(2)}
\begin{array}{rcl}
\mathbf{8} &\rightarrow& (\mathbf{4,1,1})_{1,1} \oplus (\mathbf{1,2,1})_{-2,1},\oplus (\mathbf{1,1,2})_{0,-3},
 , \\[2mm]
\end{array}
\end{equation}

We split the $\mathcal{N}=8$ index as $A=(a_1,a_2,\alpha)$, $a_1=1,2,3,5,\quad a_2=4,6,\quad \alpha=7,8,$ corresponding to the three representations in accordance with the above branching of the $\mathbf{8}$.
In this case, $W^{(\mathrm{bad})}$ reads:
\begin{align}
    W^{(\mathrm{bad})}=&[{\bf (1,1,1)}\oplus {\bf (1,1,3)}]_{(4,-2)}\oplus [{\bf (1,1,1)}\oplus {\bf (1,1,3)}]_{(0,-6)}\oplus 2\times {\bf (1,2,2)}_{(2,2)}\nonumber\\
    &\oplus [{\bf (4,1,2)}\oplus {\bf (\bar{4},2,1)}]_{(5,2)}\oplus[{\bf (\bar{4},2,1+3)}\oplus 2\times {\bf ({4},1,2)}]_{(1,-2)}\nonumber\\
    &\oplus [{\bf (4,2,1)}\oplus {\bf (\bar{4},1,2)}]_{(3,6)}\oplus [{\bf (4,2,1)}\oplus {\bf (\bar{4},1,2)}]_{(3,-6)}\oplus{\bf (6,1,1+3)}_{(2,2)}\nonumber\\
    &\oplus  {\bf (6,2,2)}_{(4,-2)}\oplus {\bf (6,2,2)}_{(0,-6)}+c.c
\end{align}

Note that, in this splitting, all non-vanishing components of $A_1$ transforming in non-singlet representations of $G_S$ belong to good representations, while for the $A_2$ tensor, and for generic $\chi$ and $\delta$, one finds a bad component transforming as:
\begin{equation}
(\mathbf{4},\mathbf{1},\mathbf{2})_{(5,2)}
\left[ A_{a_1}{}^{a_2 b_2 \alpha} \right]\,\subset \,W^{(\mathrm{bad})}
\end{equation}

However, requiring $W^{(\mathrm{bad})} = 0$ implies 
$$\delta=1\,.$$

Imposing this condition and re-evaluating the surviving components of $A_2$, one finds that the $t$-model defines a consistent truncation, capturing only the $\mathcal{N}=4$ vacuum $(\chi=0,\delta=1)$ and its $\mathcal{N}=2$ $\chi$-deformations.
\par
By considering alternative splittings, for instance $a_1=3,4,5,6$, $a_2=1,2$ and $\alpha, \beta=7,8$. of the indices for the group action ${\rm SU}(4)_S\times {\rm SU}(2)_S \times {\rm SU}(2)_R$, the resulting situation is actually more restrictive. In particular, for generic $\chi$ and $\delta$, the $A_2$ tensor contains the bad component

$$(\mathbf{6},\mathbf{2},\mathbf{2})_{(-4,2)}
\left[ A_{a_2}{}^{b_1 c_1 \alpha} \right] $$

which remains bad for any values of $\chi$ and $\delta$. The fact that now we do not even find the previous solution $\delta=1,\,\chi=0$ is due to the fact that in this limit the ${\rm SU}(4)_S$ does not reduce the structure group of the pure $\mathcal{N}=4$.

\subsection{The \texorpdfstring{$st^2$}{st\^{}2}-model truncation}
The model describes $\mathcal{N}=2$ supergravity coupled to two vector multiplets whose scalar fields span the following manifold
$$\mathcal{M}_{\rm scal.}=\left(\frac{{\rm SL}(2,\mathbb{R})}{{\rm SO}(2)}\right)^2\,.$$ 
It  is part of a consistent truncation of the maximal ungauged theory within which the above manifold is completed by a dimension-5 quaternionic-K\"ahler manifold as follows:
\begin{equation}
\left(\frac{{\rm SL}(2,\mathbb{R})}{{\rm SO}(2)}\right)^2\times \frac{{\rm SO}(4,5)}{{\rm SO}(4)\times {\rm SO}(5)}\,,
\end{equation}
From the isotropy group of the above quaternionc manifold, we can read off the structure group defining the truncation:
$${\rm SO}(4)\times {\rm SO}(5)={\rm SU}(2)_R\times {\rm SU}(2)_S\times {\rm 
USp}(4)_S\,\,\Rightarrow\,\,\,\,G_S={\rm 
USp}(4)_S\times {\rm SU}(2)_S\,.$$
The branchings of the relevant ${\rm SU}(8)$ representations with respect to $G_S\times {\rm SU}(2)_R\times {\rm U}(1)^2$ are obtained from those of the $t$-model, by branching the ${\rm SU}(4)$ representations with respect to ${\rm USp}(4)$. With respect to the $t$-model, the space $K_G$ is reduced, and thus $W^{({\rm bad})}$ augmented, by the representations in the branching of $({\bf 28}\oplus \bar{{\bf 28}})\times {\bf (5,1,1)}_{(0,0)}$.\par
For generic $\chi,\,\delta$, and different splittings on indices,  $A_2$ has the following components  in $W^{(\mathrm{bad})}$:
\begin{align}
(\mathbf{5},\mathbf{2},\mathbf{2})_{(2,2)}
\left[ A_{a_1}{}^{b_1 c_2 \alpha} \right]&\oplus (\mathbf{5},\mathbf{2},\mathbf{2})_{(0,6)}
\left[ A_{a_1}{}^{b_1 c_3 \alpha} \right] \oplus \mathbf{(5,2,2)}_{(-2,-2)} \left[A_{a_2}{}^{b_1 c_1 d_2}\right]\oplus \nonumber\\
&\oplus \mathbf{(5,2,2)}_{(-4,2)}\,\left[ A_{a_2}{}^{b_1 c_1 d_3} \right]\subset W^{(\mathrm{bad})}.
\end{align}
which never vanish for any $\chi,\delta$.

\subsection{The stu model}
This model describes $\mathcal{N}=2$ supergravity coupled to 3 vector multiplets, whose scalars span the manifold:
\begin{equation}
    \mathcal{M}_{\rm scal}=\left(\frac{{\rm SL}(2,\mathbb{R})}{{\rm SO}(2)}\right)^3\,.
\end{equation}
It  is part of a consistent truncation of the ungauged maximal theory in which the above manifold is completed by a dimension-4 quaternionic-K\"ahler space as follows \cite{Andrianopoli:1997wi}:
$$\left(\frac{{\rm SL}(2,\mathbb{R})}{{\rm SO}(2)}\right)^3\times \frac{{\rm SO}(4,4)}{{\rm SO}(4)\times {\rm SO}(4) }\,.$$
Just as in the previous cases, the structure group is read off the isotropy group of the quaternionic manifold and reads:
$$G_S={\rm SU}(2)_{S_1}\times {\rm SU}(2)_{S_2}\times {\rm SU}(2)_{S_3}\,,$$
and it is contained in those of the $t,\, t^3$ and $st^2$ models. With respect to:
$$G_s\times {\rm SU}(2)_R\times {\rm U}(1)^3\,,$$
the fundamental representation of ${\rm SU}(8)$
branches as follows:
\begin{equation}
\mathbf{8} \rightarrow
(\mathbf{2},\mathbf{1},\mathbf{1},\mathbf{1})_{(1,1,1)}
\oplus
(\mathbf{1},\mathbf{2},\mathbf{1},\mathbf{1})_{(-1,1,1)}
\oplus
(\mathbf{1},\mathbf{1},\mathbf{2},\mathbf{1})_{(0,-2,1)}
\oplus
(\mathbf{1},\mathbf{1},\mathbf{1},\mathbf{2})_{(0,0,-3)} .
\end{equation}

We find the following non-vanishing components of the $A_2$ tensor in $W^{(\mathrm{bad})}$:
\begin{align}
(\mathbf{2},\mathbf{2},\mathbf{2},\mathbf{2})_{(2,2,2)}\,[A_{a_1}{}^{b_2 b_3 \alpha}]
\oplus
(\mathbf{2},\mathbf{2},\mathbf{2},\mathbf{2})_{(-2,2,2)}\,[A_{a_2}{}^{b_1 b_3 \alpha}]
\oplus
(\mathbf{2},\mathbf{2},\mathbf{2},\mathbf{2})_{(0,-4,4)}\,[A_{a_3}{}^{b_1 b_2 \alpha}]
\subset W^{(\mathrm{bad})}
\end{align}

We conclude that no consistent truncation exists in which the unwanted representations encoded in $W^{(\mathrm{bad})}$ are absent. \par
We have also considered another $\mathcal{N}=2$ model as a possible consistent truncation of the J-Fold model. It is the ``minimal coupling'' supergravity with two vector multiplets and a scalar manifold ${\rm U}(2,1)/{\rm U}(2)$ and $G_S={\rm U}(3)$. As for the previous cases, we have verified that our conditions for consistency of truncation are not satisfied.
\section{Conclusions}
The outcome of the present work is fourfold. We have completed the analysis of \cite{Guarino:2024gke} by providing the explicit uplift formulae for the pure $\mathcal{N}'=4$ truncation of the J-fold model, in type IIB supergravity. We have proven a general condition for characterising the consistency of a truncation of a lower-dimensional supergravity which is, in turn, a consistent truncation of a maximal ten or eleven-dimensional supergravity and extended this proof to theories which do not admit an uplift. The uplift formulae were applied to the 2-charge, magnetic, non-rotating spindle solution of \cite{Ferrero:2021ovq}, thus obtaining a spindle J-fold background of the type IIB theory.
The regularity of these solutions was studied by applying a general, novel rationale for assessing the singularity structure in the uplift of an orbifold, which we give here as one of our main results.

A possible subject of a future investigation is the holographic study of the spindle J-fold background. This will involve, as the dual SCFT at the boundary, the so-called S-fold SCFT \cite{Assel:2018vtq,Bobev:2021yya}. The latter is a 2+1 theory on a superconformal interface within  $\mathcal{N}=4$ ${\rm U}(N)$-SYM and was characterized as the IR-limit of a configuration in which the SYM theory is compactified on a circle and is coupled to a $T[{\rm U}(N)]$-theory \cite{Gaiotto:2008ak}, so that the IR ${\rm U}(N) \times {\rm U}(N)$ flavour symmetry of the latter is gauged by the $\mathcal{N}=4$ vectors and a level–$n$ CS term is added, the integer $n$ being related to the monodromy along $S^1$. The peculiarity of this SCFT lies in its being dual to a non-geometric string background. As was done for the M-theory uplift of this solution on $S^7$, this solution should capture the spindle index of the gauged T[U(N)] theory \cite{Colombo:2024mts}. Using localisation techniques, this would provide a generalisation of the sphere free energies and topologically twisted indices already computed for this theory \cite{Coccia:2020wtk}, giving us insights into a non-geometric background of string theory. Another interesting task to pursue is the type IIB uplift, using the formulae presented here, for the four-dimensional solitonic solutions of \cite{Anabalon:2021tua}. We leave this to a future endeavour.

\section*{Acknowledgements}
The authors wish to thank Adolfo Guarino for his contribution in the early stages of the work and for enlightening discussions.\par
A.R. acknowledges financial support from the Department of Science, Technology, and Innovation (DSTI), South Africa, through the COST Action funding programme ‘CA22113, Fundamental Challenges in Theoretical Physics,’ and sincerely thanks the Supergravity Group in the Department of Applied Science and Technology (DISAT), Politecnico di Torino, Italy, for their hospitality and partial support during this work. A.R. and M.T. wish to thank K. Goldstein for useful discussions. C.S. is supported by a Postdoctoral Research Fellowship granted by the F.R.S.-FNRS (Belgium).
\newpage

\appendix

\section{Supersymmetric consistent truncations}
\label{App:susyCT}
In this Appendix we write, for the maximal supergravities in four, five and six dimensions, the consistency conditions for the truncation to pure $\mathcal{N}'$-extended supergravities and show that they are satisfied in the presence of a Minkowski or anti-de Sitter vacuum preserving $\mathcal{N}'$ supersymmetries.\footnote{
For a general overview of supergravities in diverse dimensions, see for instance, \cite{Andrianopoli:1996ve,Sezgin:2023hkc}.} For each truncation we define the R-symmetry and structure groups ${G}_R$, ${G}_S$, the representation of the generalized torsion tensor $W={\bf R}_\Theta$ with respect to $G_R\times G_S$, in which the fermion-shift tensors $A_1,\,A_2$ transform, and  its non-intrinsic component $W_{G_S}=\tau(K_{G_S})=K_{G_S}\cap {\bf R}_\Theta$. We then construct the non-$G_S$-singlet part of the intrinsic torsion, to be denoted by $W^{({\rm bad})}$:
$$W^{({\rm bad})}=\{{\bf R}\in W/W_{G_S}\, \vert \,{\bf R}\,\,\mbox{not singlet under } G_S\}\,.$$
Finally we show that, on a maximally symmetric vacuum preserving $\mathcal{N}'$ supersymmetries, the representations in $W^{({\rm bad})}$ are set to zero, namely that the intrinsic torsion is a $G_S$-singlet and the truncation is thus consistent.
\par The analysis below is purely group-theoretical, being based on the linear representations of the fermion-shift tensors with respect to the relevant subgroups of the R-symmetry one.
As such, we do not address the question of whether the vacua preserving  $\mathcal{N}'$ supersymmetries actually exist.

\subsection{Six dimensions}
In six dimensions, the maximal $\mathcal{N}=(2,2)$ supergravity has global symmetry group ${\rm SO}(5,5)$ and R-symmetry group ${\rm USp}(4)\times {\rm USp}(4)$.
The representations ${\rm R}_p$ of the $p$-forms are:
\begin{equation}
    {\bf R}_1={{\bf 16}}_c\,,\,\,{\bf R}_2={{\bf 10}}\,,\,\,{\bf R}_3={{\bf 16}}_s\,,\,\,{\bf R}_4={\bf 45}\,,\,\,\,{\bf R}_5={\bf R}_\Theta={\bf 144}_s\,.
\end{equation}
The gravitinos transform in the following representations of the R-symmetry group:
$$\Psi_{A\mu}\,\in \,{\bf (4,1)}\,,\,\,\,\Psi_{\dot{A}\mu}\,\in \,{\bf (1,4)}\,\,,\,\,\,\,A,\,\dot{A}=1,\dots, 4\,.$$
while the representations of the spin$-1/2$ fields are:
$$\chi_{I\,\dot{A}}=\chi_{[BC]\,\dot{A}}\in {\bf (5,4)}\,,\,\,\chi_{\dot{I}\,{A}}=\chi_{[\dot{B}\dot{C}]\,{A}}\in {\bf (4,5)}\,.$$
The fermion-shift tensors are
\begin{equation}
A_{1\,A\dot{A}}\,\in\,{\bf (4,4)}\,\,,\,\,\,\,\,A_{2\,[AB]\dot{D};\,C}\,\in\,{\bf (16,4)}\,,\,\, A_{2\,[ \dot{A}\dot{B}]D;\,\dot{C}}\,\in\,{\bf (4,16)}\,.
\end{equation}
and the embedding tensor representation ${\bf R}_\Theta$ branches, under the R-symmetry group, accordingly:
$${\bf 144}_s\,\rightarrow\,{\bf (4,4)}\oplus {\bf (16,4)}\oplus {\bf (4,16)}\,.$$
Let us consider the truncations to the following lower-supersymmetric pure supergravities.
\paragraph{Half-maximal chiral $\mathcal{N}'=(2,0)$ supergravity.} The corresponding R-symmetry and structure groups read:
$${\rm USp}(4)_R\times {\rm USp}(4)_S\,.$$
The non-intrinsic part of the torsion has the following representation content:
\begin{equation}
K_{G_S}={\bf 16}\times ({\bf 1},\,\mathfrak{g}_S)\rightarrow {\bf (4,4) }\oplus {\bf (4,20) }\oplus {\bf (4,16) }\,.
\end{equation}
In this case, the non-singlet representations with respect to ${\rm USp}(4)_S$, which belong to ${\bf R}_\Theta$ and are not in $K_{G_S}$ are:
\begin{equation}
    W^{({\rm bad})}={\bf (16,4)}\,.
\end{equation}
On a vacuum preserving $\mathcal{N}'$, the Killing spinor equations would imply:
$$A_{2\,[AB]\dot{D};\,C}=0\,,$$
setting $W^{({\rm bad})}=0$. The corresponding pure supergravity truncation is then consistent. 
\paragraph{Half-maximal $\mathcal{N}'=(1,1)$ supergravity.} The corresponding R-symmetry and structure group read:
\begin{equation}
{\rm USp}(2)_R\times {\rm USp}(2)_S\times{\rm USp}(2)_R\times {\rm USp}(2)_S\,.\label{RSgroup}\end{equation}
The structure group is $G_S= {\rm USp}(2)_S^2$ while the R-symmetry one is ${\rm USp}(2)_R^2$. Proceeding along the same lines as for the previous case, we find, with respect to the group \eqref{RSgroup}:
\begin{align}
    W^{({\rm bad})}=&{\bf (1,2,1,2)}\oplus 2\times {\bf (1,2,2,1)}\oplus 2\times {\bf (2,1,1,2)}\oplus {\bf (1,2,3,2)}\oplus {\bf (2,1,3,2)}\oplus\nonumber\\
    &\oplus{\bf (3,2,1,2)}\oplus {\bf (3,2,2,1)}\,.
\end{align}
The preserved supersymmetries are parametrized by $a_2=3,4$ and $\dot{a}_2=3,4$ while the broken ones by $a_1=1,2$ and $\dot{a}_1=1,2$, where we have split $A=\{a_1,\,a_2\}$ and $\dot{A}=\{\dot{a}_1,\,\dot{a}_2\}$.
On the $\mathcal{N}'$-supersymmetric vacuum we have the following vanishing components of the spin-$1/2$ fermion-shift:
\begin{align}
  0&=  A_{2\,[b_1{c}_2] \dot{A};\,a_2}\in {\bf (3,2,1,2)}\oplus {\bf (3,2,2,1)}\oplus {\bf (1,2,1,2)}\oplus {\bf (1,2,2,1)}\,,\nonumber\\
  0&= A_{2\,[\dot{b}_1\dot{c}_2] {A};\,\dot{a}_2}\in {\bf (1,2,1,2)}\oplus {\bf (2,1,1,2)}\oplus {\bf (1,2,3,2)}\oplus {\bf (2,1,3,2)}\,,\nonumber\\
   0&=   A_{2\,[b_1{c}_1] \dot{A};\,a_2}\in {\bf (2,1,1,2)}\oplus {\bf (2,1,2,1)}\,,\nonumber\\
      0&=   A_{2\,[\dot{b}_1\dot{c}_1] {A};\,\dot{a}_2}\in {\bf (1,2,2,1)}\oplus {\bf (2,1,2,1)}\,.\nonumber
\end{align}
which set to zero the non-singlet components of the intrinsic torsion.

\paragraph{The quarter-maximal case $\mathcal{N}'=(1,0)$.} In this case the R-symmetry group is ${\rm USp}(2)_R$ while the structure one is $G_S={\rm USp}(2)_S\times {\rm USp}(4)_S$. With respect to ${\rm USp}(2)_R\times {\rm USp}(2)_S\times {\rm USp}(4)_S$ the non-$G_S$-singlet components of the intrinsic torsion are:
$$   W^{({\rm bad})}={\bf (2,1,4)}\oplus {\bf (3,2,4)}\,,$$
which are set to zero by the Killing-spinor condition for the $\mathcal{N}'$-supersymmetric vacuum:
\begin{equation}
    0=A_{2\,[b_2,c_2]\dot{A};\,a_2}\,\in\,{\bf (2,1,4)}\,,\,\,0= A_{2\,[b_1,c_2]\dot{A};\,a_2}\,\in\,{\bf (3,2,4)}\,,
\end{equation}
where $a_1=1,2,\,a_2=3,4,$ label the broken and unbroken supersymmetry in the left-sector, respectively, while all supersymmetries in the right-sector are broken.
\paragraph{The $\mathcal{N}'=(1,2)$ case.} Finally, let us consider the case with the largest amount of preserved  
supersymmetries.  In this case $G_R={\rm USp}(2)_R\times {\rm USp}(4)_R$ while $G_S={\rm USp}(2)_S$. We can use the branching of $W$ derived earlier, interpreting it now with respect to ${\rm USp}(2)_R\times {\rm USp}(2)_S\times {\rm USp}(4)_R$. As for $K_{G_S}$, it reads:
$$K_{G_S}={\bf (1,2,4)}\oplus {\bf (1,4,4)}\oplus {\bf (2,3,4)}\,.$$
From it we deduce:
$$W^{{\rm (bad)}}={\bf (1,2,4)}\oplus {\bf (1,2,16)}\oplus {\bf (3,2,4)}\,.$$
The indices labelling the preserved supersymmetries are now $a_2=3,4$ and $\dot{A}=1,\dots,\,4$ while the broken ones are labelled by $a_1=1,2$, having split $A=\{a_1,\,a_2\}$. The Killing spinor conditions on a maximally symmetric vacuum imply:
\begin{align}
    0&=A_{2\,[b_1c_1]\dot{A};\,a_2}\in {\bf (2,1,4)}\,,\,\,  0=A_{2\,[b_1c_2]\dot{A};\,a_2}\in {\bf (1+3,2,4)}
    \,,\nonumber\\
    0&=A_{2\,[\dot{A}\dot{B}]C;\,\dot{A}}\in {\bf (1,2,16)}\oplus {\bf (2,1,16)}\,.
\end{align}
The above conditions set $W^{{\rm (bad)}}$ to zero.

\subsection{Five dimensions}
Let us consider the maximally supersymmetric theory in five dimensions and prove that the truncation to a pure $\mathcal{N}$-extended supergravity capturing a vacuum which preserves  $\mathcal{N}<8$  supersymmetries, is consistent. 
The duality group is ${\rm E}_{6(6)}$ and the relevant representations ${\bf R}_p$ of the $p$-form potentials are:
\begin{equation}
    {\bf R}_1=\overline{{\bf 27}}\,,\,\,{\bf R}_2={{\bf 27}}\,,\,\,{\bf R}_3={\bf Adj}({\rm E}_{6(6)})={{\bf 78}}\,,\,\,{\bf R}_4={\bf R}_\Theta={\bf 351}\,.
\end{equation}
The R-symmetry group is ${\rm USp}(8)$ with respect to which the two fermion shift-tensors $A_{1\,ij},\,A_{2\,i}{}^{ijk}$ transform in the following representations:
\begin{equation}
    A_{1\,ij}\,\in \,\,{\bf 36}\,,\,\,\,\,A_{2\,i}{}^{ijk}\,\in \,\,{\bf 315}\,.
\end{equation}
Let us consider the cases: $\mathcal{N}=2$, $\mathcal{N}=4$ and $\mathcal{N}=6$, in which the R-symmetry group and the structure group characterizing the pure-supergravity truncations are:
\begin{align}
\mathcal{N}=2\,:\,\,\,{\rm USp}(2)_R\times {\rm USp}(6)_S\subset {\rm USp}(8)\,,\nonumber\\
\mathcal{N}=4\,:\,\,\,{\rm USp}(4)_R\times {\rm USp}(4)_S\subset {\rm USp}(8)\,,\nonumber\\
\mathcal{N}=6\,:\,\,\,{\rm USp}(6)_R\times {\rm USp}(2)_S\subset {\rm USp}(8)\,,\nonumber
\end{align}
In the three cases we split the R-symmetry index as follows:
$$i=(a_1,\,a_2)\,\,;\,\,\,a_1=1,\dots, 8-\mathcal{N}\,;\,\,a_2=9-\mathcal{N},\dots, 8\,,$$
where $a_2 $ labels the unbroken supersymmetries.

\paragraph{$\mathcal{N}=2$ case.} The branchings of the $A_1,\,A_2$ tensor representations and of the ${\bf 27}$,  with respect to ${\rm USp}(2)_R\times {\rm USp}(6)_S$ are:
\begin{align}
   {\bf 27}&\rightarrow  {\bf (1,1)}\oplus {\bf (1,5)}\oplus {\bf (5,1)}\oplus {\bf (4,4)} \nonumber\\
   {\bf 36}&\rightarrow {\bf (3,1)}\oplus {\bf (1,21)}\oplus {\bf (2,6)}\,,\nonumber\\
    {\bf 315}&\rightarrow {\bf (3,14)}\oplus {\bf (2,14')}\oplus {\bf (2,6)}\oplus {\bf (1,70)}\oplus {\bf (1,21)}\oplus {\bf (1,14)}\oplus {\bf (2,64)}\,.
\end{align}
The space $K_{G_S}$ branches as follows:
\begin{equation}
   K_{G_S}=\mathbf{27} \otimes (\mathbf{1},\,\mathrm{Adj}\,G_S)\rightarrow\,2\times \mathbf{(1,21)}\oplus {\bf (1,70)}\oplus {\bf (1,14)}\oplus {\bf (2,64)}\oplus {\bf (2,6)}\oplus {\bf (1,189)}\oplus {\bf (2,56)}\,.
\end{equation}
The bad representations are:
$$W^{{\rm (bad)}}={\bf (2,14')}\oplus{\bf (3,14)}\oplus {\bf (2,6)}\,.$$
We see that these are precisely the representations which are set to zero by the condition:
$$A_{a_2}{}^{ijk}=0\,\Leftrightarrow\,\, 0=A_{2\,a_2}{}^{b_2\,a_1 b_1}\,[{\bf (3+1,14)}]\,,\,\,\, A_{2\,a_2}{}^{b_1\,c_1 d_1}\,[{\bf (2,14')}]\,,\,\,A_{2\,a_2}{}^{b_2\,c_2 d_1}\,[{\bf (2,6)}]\,,$$
which holds at the $\mathcal{N}$-supersymmetric vacuum.

\paragraph{$\mathcal{N}=4$ case.}
\begin{align}
\mathbf{27}&\rightarrow \mathbf{(1,1)} \oplus\mathbf{(5,1)}\oplus \mathbf{(1,5)}\oplus\mathbf{(4,4)}   \,,\nonumber\\
\mathbf{36} &\rightarrow \mathbf{(10,1)} \oplus \mathbf{(1,10)} \oplus \mathbf{(4,4)}, \nonumber\\
\mathbf{315} &\rightarrow \mathbf{(16,4)} \oplus \mathbf{(4,16)} \oplus 2\times \mathbf{(4,4)} \oplus \mathbf{(10,5)} \oplus \mathbf{(10,1)} \oplus \mathbf{(5,10)} \nonumber\\
&\qquad \oplus \mathbf{(5,5)}\oplus \mathbf{(5,1)}\oplus \mathbf{(1,10)}\oplus \mathbf{(1,5)}\,.
\end{align}

The space $K_{G_S}$ branches as follows:
\begin{align}
K_{G_S} &= \mathbf{27} \otimes (\mathbf{1},\,\mathrm{Adj}\,G_S) \rightarrow\nonumber\\
&\rightarrow  \mathbf{(1, 35)} \oplus \mathbf{(1,5)} \oplus \mathbf{(5,10)} \oplus 2\times \mathbf{(1,10)}  \oplus \mathbf{(4,20)} \nonumber\\
&\quad \oplus \mathbf{(4,16)} \oplus \mathbf{(4,4)}\,.
\end{align}

The bad representations are:
$$W^{{\rm (bad)}}={\bf (16,4)}\oplus 2\times  {\bf (4,4)}\oplus {\bf (5,5)}\oplus {\bf (10,5)}\,.$$
$\mathcal{N}=4$ supersymmetry
sets the following representations to zero:
$$2\times {\bf (4,4)}\oplus {\bf (1\oplus 10\oplus5,5)}\oplus {\bf (16,4)}\,,$$
which sets $W^{{\rm (bad)}}$ to zero.
Indeed, we have:
$$A_{a_2}{}^{ijk}=0\,\Leftrightarrow\,\, 0=A_{2\,a_2}{}^{b_2\,a_1 b_1}\,[{\bf (1\oplus 5\oplus 10,5)}]\,,\,\,\, A_{2\,a_2}{}^{b_1\,c_1 d_1}\,[{\bf (4,4)}]\,,\,\,A_{2\,a_2}{}^{b_2\,c_2 d_1}\,[{\bf (4\oplus 16,4)}]\,,$$
So the pure $\mathcal{N}=4$ supergravity truncation is consistent.

\paragraph{$\mathcal{N}=6$ case.}
We can use the above branching of $W={\bf 351}$, interpreting it now with respect to ${\rm USp}(2)_S\times {\rm USp}(6)_R$. We now compute, with respect to ${\rm USp}(6)_R\times {\rm USp}(2)_S$ the non-$G_S$-singlet components of the intrinsic torsion:
$$W^{(\rm bad)}={\bf (14',2)}\oplus {\bf (6,2)}\oplus {\bf (64,2)}\,.$$
These are set to zero by the Killing spinor equation on the $\mathcal{N}'=6
$-supersymmetric vacuum, which implies:
$$0=A_{2\,a_2}{}^{b_2 c_2\,a_1}\,\in\,{\bf (14'\oplus 6\oplus 64,2)}\,.$$

\subsection{Four dimensions}

\paragraph{$\mathcal{N}=1$ case.} We write down the branching of the relevant $\SU(8)$ representations under $\SU(7) \times \Urm(1)$:
\begin{equation}
    \begin{array}{rcl}
    \mathbf{8} &\rightarrow& \mathbf{7}_{1} \oplus \mathbf{1}_{-7} \\
    \mathbf{28} &\rightarrow&\mathbf{7}_{-6}\oplus \mathbf{21}_{2} \\
    \mathbf{36} &\rightarrow& \mathbf{1}_{-14}\oplus\mathbf{7}_{-6}\oplus\mathbf{28}_{2}\\
    \mathbf{420} &\rightarrow&  \mathbf{21}_{2}\oplus\mathbf{35}_{10}\oplus\mathbf{140}_{-6}\oplus\mathbf{224}_{2}\\
    \end{array}
\end{equation}
Then we compute the various irreducible representations present in the space of compatible connections $K_{\SU(7)} = \mathbf{56} \otimes (\mathbf{49}_0)$:
\begin{equation}
\begin{split}
    K_{\SU(7)}  =&  \mathbf{7}_{-6} \oplus\mathbf{140}_{-6}  \oplus\mathbf{21}_{2} \oplus\mathbf{28}_{2} \oplus\mathbf{224}_{2} \oplus \,\text{c.c.} 
\end{split}
\end{equation}
Thus consistency imposes that the following representation must vanish from the A-matrices at the $\mathcal{N}=1$ vacuum:
\begin{equation}
    W^{(\text{bad})} = \mathbf{35}_{10} \oplus \,\text{c.c.}\,.
\end{equation}
The bad representation can be identified with the component $A_{\alpha}{}^{abc}$ (where $a=2,\,\dots, \,8$ labels the $\SU(7)$ representation and $\alpha =1$). Since supersymmetry imposes $A_{\alpha}{}^{\dots} = 0$, the truncation is always consistent.

Moreover, via the quadratic constraints, evaluating the first equations of (2.6) in \cite{Gallerati:2014xra} with $k=m=\alpha$, $n=a$ and $l=b$, we get that
\begin{equation}
    A^{\alpha}{}_{
    ba \gamma} A^{\beta\gamma} =0\,.
\end{equation}
Giving us the proof of consistency for $G_S= \SU(7)$ structure as in section \ref{subsec:prooflowDim} whenever $A_{\alpha\beta} \neq 0$, which is the case of interest for consistent truncations around AdS$_4$ solutions. The resulting model is the pure $\mathcal{N}=1$ supergravity with cosmological constant, originally constructed in \cite{Townsend:1977qa}.

\paragraph{$\mathcal{N}=2$ case.}
We show that the $\SU(6)$-structure induced by a given vacuum in any maximal gauged supergravity always provide a consistent truncation to pure $\mathcal{N}=2$ supergravity. We write down the branching of the relevant $\SU(8)$ representations under $\SU(6) \times \SU(2) \times \mathrm{U}(1)$:
\begin{equation}
\begin{array}{rcl}
    \mathbf{8} &\rightarrow& (\mathbf{6},\,\mathbf{1})_{-1} \oplus (\mathbf{1},\, \mathbf{2})_{3}\\
    \mathbf{28} &\rightarrow& (\mathbf{6},\,\mathbf{2})_{-2}\oplus (\mathbf{15},\,\mathbf{1})_{2}\oplus (\mathbf{1},\,\mathbf{1})_{-6}\\
    \mathbf{36} &\rightarrow& (\mathbf{21},\,\mathbf{1})_2 \oplus (\mathbf{6},\,\mathbf{2})_{-2} \oplus (\mathbf{1},\,\mathbf{3})_{-6}\\
    \mathbf{420} &\rightarrow& (\mathbf{35},\,\mathbf{1})_{-6} \oplus (\mathbf{84},\,\mathbf{2})_{-2} \oplus (\mathbf{6},\,\mathbf{2})_{-2}\oplus \\&&(\mathbf{105},\,\mathbf{1})_{2}\oplus (\mathbf{15},\,\mathbf{3})_{2}\oplus (\mathbf{15},\,\mathbf{1})_{2}\oplus (\mathbf{20},\,\mathbf{2})_{6}
\end{array}
\end{equation}
Then we compute the various irreducible representations present in the space of compatible connections $K_{\SU(6)} = \mathbf{56} \otimes (\mathbf{35},\,\mathbf{1})_0$:
\begin{equation}
\begin{split}
    K_{\SU(6)}  =&  (\mathbf{120},\,\mathbf{2})_{-2}\oplus (\mathbf{84},\,\mathbf{2})_{-2}\oplus (\mathbf{6},\,\mathbf{2})_{-2}\oplus\\
    & (\mathbf{384},\,\mathbf{1})_{2}\oplus (\mathbf{105},\,\mathbf{1})_{2}\oplus (\mathbf{21},\,\mathbf{1})_{2}\oplus (\mathbf{15},\,\mathbf{1})_{2}\oplus\\
    &  (\mathbf{35},\,\mathbf{1})_{-6}\,.
\end{split}
\end{equation}
Thus consistency imposes that the following representations must vanish from the $A$-matrices at the $\mathcal{N}=2$ vacuum:
\begin{equation}
    W^{(\text{bad})} = (\mathbf{6},\,\mathbf{2})_{-2}\oplus(\mathbf{15},\,\mathbf{3})_{2}\oplus(\mathbf{20},\,\mathbf{2})_{6} + c.c.
\end{equation}
The bad representation can be identified with the components in $A_{\alpha a}$, $A_{\alpha}{}^{\beta\gamma a}$, $A_\alpha{}^{\beta ab}$ and $A_{\alpha}{}^{abc}$ (where $\alpha=1,\,2$ labels the fundamental $\SU(2)$ representation and $a = 1,\,\dots,\,6$ labels the $\SU(6)$ representations). Since supersymmetry imposes that $A_{\alpha}{}^{\cdots} = 0 = A_{\alpha a}$, the truncation is always consistent. 

\paragraph{$\mathcal{N}=3$ case.}

We consider the truncation down to pure \(\mathcal{N}=3\) theory for which $G_S={\rm SU}(5)_S$: 
\[
\SU(8) \longrightarrow {\rm SU}(5)_S\times {\rm SU}(3)_R\times  {\rm U}(1)_R
\]

The relevant branching of ${\rm SU}(8)$-representations with respect to $G_S\times {\rm SU}(3)_R\times  {\rm U}(1)_R$ are:
\begin{equation}
\label{branching wrt SU(5)*SU(3)}
\begin{array}{rcl}
\mathbf{8} &\rightarrow& (\mathbf{5,1})_{3} \oplus (\mathbf{1,3})_{-5}, \\[2mm]
\mathbf{28} &\rightarrow&(\mathbf{10,1})_{(6)} \oplus
(\mathbf{5,3})_{(-2)} \oplus
(\mathbf{1,\bar{3}})_{(-10)}  \\[2mm]
\mathbf{36} & \rightarrow &
(\mathbf{15,1})_{(6)} \oplus (\mathbf{5,3})_{(-2)} \oplus (\mathbf{1,6})_{(-10)} ,
\\[2mm]
\mathbf{420} & \rightarrow &
(\mathbf{\bar{5},1})_{(-18)} \oplus
(\mathbf{24, \bar{3}})_{(-10)} \oplus
(\mathbf{1,\bar{3}})_{(-10)} \oplus
(\mathbf{45,3})_{(-2)} \oplus
(\mathbf{5,\bar{6}})_{(-2)} \oplus
\\&
&
(\mathbf{5,3})_{(-2)} \oplus
(\mathbf{40,1})_{(6)} 
\oplus
(\mathbf{10,8})_{(6)}\oplus
(\mathbf{10,1})_{(6)}\oplus
(\mathbf{\bar{10},\bar{3}})_{(14)}\,.

\end{array}
\end{equation}
We also find
\begin{equation}
\label{eqn: K}
\begin{array}{rcl}
       K_{G_S} 
&=& [\mathbf{28} \oplus cc] \otimes 
[(\mathbf{24,1})]_{0}
\\
&=& [\mathbf{(10,1})_{(6)} \oplus
(\mathbf{5,3})_{(-2)} \oplus
(\mathbf{1,\bar{3}})_{(-10)}]  \otimes 
[(\mathbf{24,1})
\\
&=&
\mathbf{(10 \otimes 24,1})_{(6)} \oplus
(\mathbf{5 \otimes 24,3})_{(-2)} \oplus
(\mathbf{24,\bar{3}})_{(-10)}
\\
&=&
\mathbf{(175 \oplus 40 \oplus 15 \oplus 10,1})_{(6)} \oplus
(\mathbf{70 \oplus 45 \oplus 5,3})_{(-2)} \oplus
(\mathbf{24,\bar{3}})_{(-10)}
\oplus c.c.
\end{array}
\end{equation}
From which we find the following bad representations:
\begin{equation}
    W^{({\rm bad})}= ({\bf 5,1})_{(18)}\oplus {\bf (\bar{10},\bar{3})}_{(14)}\oplus{\bf (10,8)}_{(6)}\oplus {\bf (\bar{5},6\oplus \bar{3})}_{(2)}\oplus c.c.
\end{equation}

Note that, in this splitting, the non-vanishing entries of $A_1$ are:
\begin{align}
A_{1\,ab}&\in ({\bf 15,1})_{(6)}\,,\nonumber\\
A_{1\,a\alpha}&\in ({\bf 5,3})_{(-2)}\,,\nonumber\\
A_{1\,\alpha\beta}&\in ({\bf 1,6})_{(-10)},\nonumber
\end{align}
Without loss of generality, we can set $A_{1\,a\alpha}=0$ and $A_{\alpha\beta}\propto \delta_{\alpha\beta}$ at the vacuum. The bad representation of $A_2$ can be identified with the components in 
$A_{a}{}^{\alpha\beta\gamma} \in ({\bf 5,1})_{(18)}$, $A_{\alpha}{}^{\beta\gamma a} \in ({\bf \bar{5},{6}+\bar{3}})_{(2)}$, 
$A_{\alpha}{}^{\beta ab} \in ({\bf \bar{10},8})_{(-6)}$, 
$A_{\alpha}{}^{abc}\in {\bf ({{10}},{3})}_{-14}$. The last three tensors are set to zero by the Killing spinor equations on a $\mathcal{N}'=3$-supersymmetric vacuum. The vanishing of the first representation follows from the T-identities on the supersymmetric vacuum, which imply $A_{a}{}^{\alpha\beta\gamma}\,A_{\gamma\delta}=0$ \cite{Gallerati:2014xra}. 

(where $\alpha=1,\,2 ,\,3$ labels the fundamental $\SU(3)$ representation corresponds to the preserved supersymmetries and $a = 1,\,\dots,\,5$ labels the $\SU(5)$ representations corresponds to the broken supersymmetries).

In the J-fold gauging, on the vacua parametrised by $\delta,\,\chi$, we always find, for generic parameters, two non-vanishing bad representations:
$$A_{\alpha}{}^{abc}\in {\bf ({{10}},
{3})}_{-14}\,,\,\,\,A_{\alpha}{}^{\beta ab}\in  {\bf (\bar{10},{8})}_{-6}\,,$$
which are zero only if the 
 3 supersymmetries are chosen among the preserved $\mathcal{N}=4$ ones and  $\delta=1,\,\chi=0$. This implies that the only consistent pure $\mathcal{N}=3$ truncation is a truncation of the pure $\mathcal{N}=4$, with gauging ${\rm SO}(3)$.

\section{Orbifolds and regularity of uplifts}
\label{app:orbifolds}

In this appendix we will introduce several notions related to orbifolds and $G$-orbibundles. Most details presented here are standard results and are not necessary to apply the result of section \ref{sec:Regularity}. However, some details and applications to the case where $G$ is abelian are useful to further characterise the types of singularities of the uplifted solution. We refer to \cite{Sparks:2010sn} and references therein for more details on the subject.
\subsection{Orbifolds and orbibundles}
We recall some notions for orbifolds keeping as much as possible the discussion straightforward and relevant for the physics described in the paper.

\paragraph{Definitions} Like manifolds, an orbifold $\mathcal{O}$ is a topological space $|\mathcal{O}|$ endowed with an equivalency class of atlases, themselves built from a set of charts. These charts are given by a triple $(\tilde{U_i},\,\phi_i,\,\Gamma_i)$ where 
\begin{itemize}
    \item $U_i \subset \mathbb{R}^n$ is an open set of $\mathbb{R}^n$;
    \item $\Gamma_i$ is a discrete subgroup of $\GL(n)$ acting faithfully on $\tilde{U}_i$;
    \item $\phi_i: \tilde{U}_i \rightarrow |\mathcal{O}|$ are $\Gamma_i$-invariant continuous maps. 
\end{itemize}
We require the $U_i:=\phi_i(\tilde{U_i})$ to provide an open cover of $|\mathcal{O}|$ as well as compatibility conditions between the various charts whenever their images overlap, yielding to a notion of smoothness on orbifold similar to that of manifolds.

\paragraph{Singularities and isotropy group} We stress that there is no notion of orbifold ``singularity" for the underlying topological space $|\mathcal{O}|$. This notion is captured by the extra structure on $\mathcal{O}$ and the notion of isotropy groups of points on $|\mathcal{O}|$. Let $(\tilde{U}_i,\,\phi_i,\,\Gamma_i)$ be a chart containing a point $o \in |\mathcal{O}|$, i.e. $o\in \phi_i(\tilde{U}_i)$, the isotropy group of $o$ is $\Gamma_o = \text{Stab}_{\phi^{-1}(o)} \subset \Gamma_i$. This group is independent of the choice of chart and characterises orbifold singularities at $o$. We say that the orbifold is regular at $o$ if $\Gamma_o = \left\{e\right\}$.

\paragraph{Function between orbifolds}

In the same way that smooth map between smooth manifolds are those preserving the differentiable structure given by the charts, smooth maps between orbifolds must be compatible with the extra data given by the orbifold charts. 

Let $\mathcal{O}_1$ and $\mathcal{O}_2$ be two orbifolds, an orbifold map $f:\mathcal{O}_1\rightarrow \mathcal{O}_2$ is given by a map $|f|:|\mathcal{O}_1|\rightarrow |\mathcal{O}_2|$ as well as a series of lift for the local models: for local charts $(\tilde{U}_p,\,\phi_p,\,\Gamma_p)$ and $(\tilde{U}_{f(p)},\,\phi_{f(p)},\,\Gamma_{f(p)})$, one must specify a lift $\tilde{f}_p: \tilde{U}_p \rightarrow \tilde{U}_{f(p)}$ such that 

\begin{equation}
\label{eq:commDiagMapOrbifold}
\begin{tikzcd}
	{\tilde{U}_p} & {\tilde{U}_{f(p)}} \\
	{U_p} & {U_{f(p)}}
	\arrow["{\tilde{f}}", from=1-1, to=1-2]
	\arrow["{\phi_p}"', from=1-1, to=2-1]
	\arrow["{\phi_{f(p)}}", from=1-2, to=2-2]
	\arrow["{|f|}"', from=2-1, to=2-2]
\end{tikzcd}\end{equation}
commutes. This induces a morphism $f_{\Gamma_p}:\Gamma_p \rightarrow \Gamma_{f(p)}$ under which $\tilde{f}$ must be equivariant.

An example which will prove useful is the diagonal map $\Delta : \mathcal{O} \rightarrow \mathcal{O}\times \mathcal{O}:o\mapsto (o,\,o)$. For each $o \in \mathcal{O}$, we further define the local lifts $\tilde{f}: \tilde{U}_o \rightarrow  \tilde{U}_o \times  \tilde{U}_o$. One can check that \eqref{eq:commDiagMapOrbifold} does commute with $\tilde{f}$ which is equivariant under the induced morphism $f_\Gamma:\Gamma_p \rightarrow \Gamma_p \times \Gamma_p: g\mapsto (g,\,g)$.

\paragraph{Orbibundles}
Orbibundle are the orbifold generalisation of fibre bundles. In particular, given a Lie group $G$, a principal orbibundle is a smooth manifold\footnote{Orbibundles are sometimes allowed to be orbifolds themselves. Here, we restrict the definition and require that the total space $P$ is a smooth manifold} $P$ which admits a \emph{locally free} action of $G$ and local trivialisations. By definition of a locally free action, the stabiliser of a point $p\in P$ is a discrete group $\Gamma_p \in G$. This endows the base space $\mathcal{O}=P/G$ with an orbifold structure whose isotropy group at a point $o$ is $\Gamma_o = \Gamma_{p \in \pi^{-1}(o)}$ (the groups $\Gamma_p$ are all equivalent up to $G$ conjugation for $p \in \pi^{-1}(o)$). Thus the local trivialisations are of the form $U_o \times_{\Gamma_p} G$ where $U_o$ are the local covering space of neighbourhood of points in $\mathcal{O}$.

\paragraph{Pullbacks}
As usual for fibre bundles, we can consider the pullback of an orbibundle by an orbifold map. Let $f:\mathcal{O}'\rightarrow \mathcal{O}$ be an orbifold map and $P$ a $G$-orbibundle on $\mathcal{O}$. We define the pullback of $P$ by $f$ as
\begin{equation}
    f^* P = \{(o',\,p) \in O'\times P\,|\,f(o') = \pi(p)\}.
\end{equation}
The orbibundle projection is defined by the projection on the first factor. Moreover, $f^*P$ still admits a locally free $G$-action given by the action on the second factor. Finally, $f^*P$ is still a smooth manifold. Indeed, in any local trivialisation, the lift of $\tilde{f}$ provides a pull-back of the trivial bundle $U_p\times G$ as $\tilde{f}^*(U_p \times G)$. Since this function is $\Gamma_p$ equivariant, we can pass through the quotient $\tilde{f}^*(U_p \times G)/\Gamma_{f(p)} = f^*(U_p \times_{\Gamma_p} G)$ and smoothness is obtained by realising that $\Gamma_{f(p)}$ acts freely on $\tilde{f}^*(U_p \times G)$. This definition provides embeddings $f_\Gamma:\Gamma_{f(p)} \rightarrow G$ inherited from the embeddings $\Gamma_p \rightarrow G$ of the original orbibundle. 

In particular, we can use this construction to show that given two principal orbibundle $P_1$ and $P_2$ with structure groups $G_1$ and $G_2$ over the same orbifold $O$, their product
\begin{equation}
    P_1 \times_O P_2 := \Delta^*(P_1 \times P_2)
\end{equation}
is a $G_1 \times G_2$ orbibundle whose total space is smooth, and the embedding $\Delta_\Gamma : \Gamma_p \rightarrow G_1 \times G_2$ is the diagonal embedding.

\subsection{Regularity of uplifts}
Let us consider an orbifold solution of a lower dimensional gauged supergravity with compact gauge group $G$. Various quantisation conditions must be imposed on the lower-dimensional gauge fields $A_\mu$, allowing us to consider them as a connection on a principal orbibundle $P$. The uplift of this solution, using consistent truncation techniques, will be of the form
\begin{equation}
    \Mtot = P\times_G \Mint\,,
\end{equation}
where $\Mint$ must admit an effective $G$ action \cite{Rovere:2025jks}. By the slice theorem \cite{Meinrenken2003}, any point $m\in\Mint$ admits an equivariant local neighbourhood of the form 
\begin{equation}
    U_m = G\times_{G_m} V_m
\end{equation}
where $G_m$ is the stabiliser of $m$ and $V_m = T_m\Mint/T(G\cdot m)$ admits a $G_m$ action. Importantly, both $G_m$ and $V_m$ may vary depending on the choice of point in $\Mint$. Now, using a local trivialisation $(U_p,\,\Gamma_p)$ of $P$, we have the equivariant local neighbourhood
\begin{equation}
    U_{(p,\,m)} =\left(U_p \times_{\Gamma_p}\times G\right)\times \left(G \times_{G_m} V_m\right)\,.
\end{equation}
and 
\begin{equation}
\begin{split}
    U_{[p,\,m]} =&\left(U_p \times_{\Gamma_p} G\right)\times_G \left(G \times_{G_m} V_m\right)
\end{split}
\end{equation}
which is smooth whenever the $G$ action on $U_{(p,\,m)}$ is free. If not, we can extract the type and position of singularities from $\Gamma_p$ and $G_m$.

Let us apply this to the type IIB uplift of the spindle solution. In that case, $G=\Urm(1)^2$, $\Gamma_{p}$ is non-trivial only at the orbifold point of the spindle where it is isomorphic to $\mathbb{Z}_{n_\pm}$ and embeds in $\Urm(1)^2$ as 
\begin{equation}
    \mathbb{Z}_{n_\pm} \rightarrow \Urm(1)^2 : \omega_{n_\pm} \rightarrow (\omega^{p_1}_{n_\pm},\,\omega^{p_2}_{n_\pm})\,.
\end{equation} 
The internal manifold is of the form
\begin{equation}
    S^5 \times S^1 \cong S^2 \times S^2 \times \Sigma\,,
\end{equation}
and the U(1) action, acting as rotations of the sphere around their poles, is not free only at the poles of the two-spheres. All points, except from those in the preimage of the spindle singularities, are regular so we will only consider the uplift near the orbifold points. We thus have the following local model:
\begin{equation}
    V_m = \mathbb{C}_1 \times \mathbb{C}_2 \times \mathbb{R}^2_{\Sigma}\hspace{5mm},\hspace{5mm}
G_m = \Urm(1)^2\,,
\end{equation}
and the space is locally
\begin{equation}
    U_{[p,\,m]} = AdS_2 \times \frac{\mathbb{C} \times \mathbb{C}_1\times \mathbb{C}_2}{\mathbb{Z}_{n_\pm}}\times \mathbb{R}^2_\Sigma\,.
\end{equation}
The action of the isotropy group, focus on $\mathbb{Z}_{n_-}$ for concreteness, is
\begin{equation}
    \omega_{n_-}\cdot (z,\,z_1,\,z_2) = (\omega^{n_+}_{n_-}\cdot z,\,\omega^{p_1}_{n_-} \cdot z_1,\,\omega^{p_2}_{n_-} \cdot z_2)\,,
\end{equation}
which is singular at $z = z_i = 0$. We can render this action more familiar by using another generator for $\mathbb{Z}_{n_-}$, $\omega_{n_-}^{m_-}$, such that $\omega_{n_-}^{m_- n_+} =\omega_{n_-} $. This is equivalent to requiring
\begin{equation}
    m_- = n_+^{-1}\,\text{mod}\,n_- \hspace{5mm}\Leftrightarrow \hspace{5mm} m_- n_+ - m_+ n_-=1
\end{equation}
for which a solution always exists if $gcd(n_+,\,n_-) = 1$ (Bézout-Euclid). We finally obtain the action of $\mathbb{Z}_{n_-}$ near the poles to be
\begin{equation}
    \omega_{n_-} (z,\,z_1,\,z_2) = (\omega_{n_-} z,\,\,\omega^{m_-p_1}_{n_-} \cdot z_1,\,\omega^{m_-p_2}_{n_-} \cdot z_2)\,.
\end{equation}
Equivalent results can be obtained for the south poles of the spheres where $p_i \rightarrow -p_i$ and at the other tip of the spindle with $\pm \rightarrow \mp$.

We insist that those results are entirely obtained by considering the $G$ action on $\Mint$, or more precisely, by considering the stabilisers $G_m$ of points in $\Mint$. No further knowledge of the solution is necessary when it comes to orbifold singularities.

\bibliography{Draft}

\end{document}